\documentclass[a4paper]{nature}
\usepackage[utf8]{inputenc}
\UseRawInputEncoding
\usepackage[english]{babel}
\usepackage[]{graphicx}
\usepackage{amsmath}
\usepackage{amssymb}
\usepackage{mathrsfs}
\usepackage{textcomp}
\usepackage{tabularx}
\usepackage{color}
\usepackage{array}
\usepackage{float}
\usepackage{braket}
\usepackage{mathtools}
\restylefloat{table}

\definecolor{darkblue}{rgb}{0.1,0.2,0.6}
\definecolor{darkred}{rgb}{0.8,0.1,0.2}
\usepackage[colorlinks,citecolor=darkblue,linkcolor=darkred,urlcolor=darkblue]{hyperref}

\makeatletter
\let\saved@includegraphics\includegraphics
\AtBeginDocument{\let\includegraphics\saved@includegraphics}
\renewenvironment*{figure}{\@float{figure}}{\end@float}
\makeatother

\title{Strongly anharmonic flux-tunable transmon based on InAs-Al 2D heterostructure}

\author{Shukai Liu$^{1,^{*},^{\dagger}}$, Arunav Bordoloi$^{1,2,^{*},^{\dagger}}$, Jacob Issokson$^{2}$, Ido Levy$^{2}$, Maxim G. Vavilov$^{3}$, Javad Shabani$^{2}$ \& Vladimir E. Manucharyan$^{1,4}$}

\begin{document}

\maketitle
\thispagestyle{empty}

\begin{affiliations}
	\item Department of Physics, Joint Quantum Institute, and Quantum Materials Center, University of Maryland, College Park, MD, USA
	\item Center for Quantum Information Physics, Department of Physics, New York University, New York, NY, USA
	\item Department of Physics, University of Wisconsin-Madison, Madison, WI, USA
        \item Institute of Physics, Ecole Polytechnique Federale de Lausanne, Lausanne, Switzerland
	\item[$^{\dagger}$] These authors contributed equally to this work.
        \item[$^{*}$] Corresponding authors: ab13024@nyu.edu, sliu499@umd.edu
\end{affiliations}

\begin{abstract}
The gatemon qubits, made of transparent superconducting-semiconducting Josephson junctions, typically have even weaker anharmonicity than the opaque AlOx-junction transmons. However, flux-frustrated gatemons can acquire a much stronger anharmonicity, originating from the interference of the higher-order harmonics of the supercurrent. Here we investigate this effect of enhanced anharmonicity in split-junction gatemon devices based on InAs-Al 2D heterostructure. We find that anharmonicity in excess of 100\% can be routinely achieved at the half-integer flux sweet-spot without any need for electrical gating or excessive sensitivity to the offset charge noise. We verified that such intrinsically large anharmonicity enables our devices to be driven coherently with raw Rabi frequencies exceeding 100 MHz, without any pulse shaping, simplifying implementation and control compared to traditional gatemons and transmons. Furthermore, by analyzing a relatively high-resolution spectroscopy of the device transitions as a function of flux, we were able to extract fine details of the current-phase relation, to which transport measurements would hardly be sensitive.  The strong anharmonicity of our anharmonic tunable transmons, along with their bare-bones design, can prove to be a precious resource that transparent superconducting-semiconducting junctions bring to quantum information processing.

\end{abstract}

\maketitle


\textit{Introduction.} Electrically-gated transmon qubits (gatemons) have been demonstrated as potentially useful devices for building large scale quantum processors~\cite{casparis_gatemon_2016, de_lange_realization_2015, sagi_gate_2024,hertel_gate-tunable_2022,luthi_evolution_2018, larsen_semiconductor-nanowire-based_2015, huo_gatemon_2023, zhuo_hole-type_2023, zheng_coherent_2024, danilenko_few-mode_2023, erlandsson_parity_2023, strickland_gatemonium_2024, pita-vidal_gate-tunable_2020, casparis_superconducting_2018, kiyooka_gatemon_2025}. A typical gatemon device is made with a gate electrode close to a hybrid superconductor-semiconductor Josephson junction (superconducting-semiconducting JJ). The voltage on the gate electrode changes the electron density in the semiconductor weak link and thereby tunes the Josephson energy $E_{\textrm{J}}$ of the qubit. The advantage of electrical gating is that it can be used for controlling the quantum dynamics by rapidly changing the qubit frequency, similarly to switching transistors in conventional semiconductor circuits. Just like transmons, however, gatemons suffer from a suppressed anharmonicity of the qubit transition, as they operate in a regime where the Josephson phase-difference is localized near the bottom of the Josephson potential well \cite{koch_charge-insensitive_2007, houck_life_2009, schreier_suppressing_2008, kringhoj_suppressed_2020, bargerbos_observation_2020}. Increasing the anharmonicity, which we define by
\begin{equation}
\label{eq:anharm}
\eta = \left|\frac{f_{12}-f_{01}}{f_{01}}\right|,
\end{equation} 
would require increasing the charging energy $E_{\textrm{C}}$, which would exponentially increase the qubit's sensitivity to the offset-charge noise. Furthermore, the relatively high transparency of superconducting-semiconducting junctions further reduces anharmonicity in comparison to opaque AlOx transmon junctions~\cite{kringhoj_anharmonicity_2018}. The weak anharmonicity of gatemons would limit gatemon-based quantum processors in the same way it limits transmons, that is single-qubit and two-qubit gate operations will have to be done sufficiently slowly as to prevent the leakage outside the computational space \cite{steffen_accurate_2003, chow_randomized_2009}.

Replacing the gatemon junction with a split junction and applying a flux bias through the loop can dramatically increase the qubit anharmonicity in comparison to an equivalent split-junction transmon device \cite{de_lange_realization_2015, larsen_parity-protected_2020}. This effect originates from a destructive interference of the first Josephson harmonics and a constructive interference of the typically much weaker second Josephson harmonics, bringing their contribution on par \cite{ciaccia_charge-4e_2024, hart_current-phase_2019, zhang_large_2024, banszerus_hybrid_2024, banszerus_voltage-controlled_2024, leblanc_gate-_2025, fominov_asymmetric_2022}.  As a result, the Josephson potential energy, while remaining a $2\pi$-periodic function, acquires a more complex shape involving local minima, which can lead to much more anharmonic spectra. By contrast, higher harmonics are suppressed in conventional tunnel junctions, and hence flux bias merely rescales the regular Josephson potential, with weak effect on the anharmonicity \cite{houck_controlling_2008}. So far, higher harmonics of the supercurrent has been mainly interested in the context of parity-protected qubits, which are possible when the first harmonic is completely canceled. In practice, though, such cancellation requires a precise control of either the junction parameters or the electrical gating, and in practice such protected qubits remain a distant goal \cite{larsen_parity-protected_2020, ciaccia_charge-4e_2024, banszerus_hybrid_2024}. In this paper, we describe a more immediate impact of the supercurrent interference in two transparent junctions by focusing on the half-flux quantum sweet spot, where the anharmonicity is already dramatically enhanced but the sensitivity to the 1/f flux noise is zero to first order. Such a device presents an opportunity to improve on the main drawback of transmons - the weak anharmonicity -- using the unique property of mesoscopic superconducting-semiconducting junctions.
\begin{figure*}[]
	\centering
	\includegraphics[width=1\textwidth]{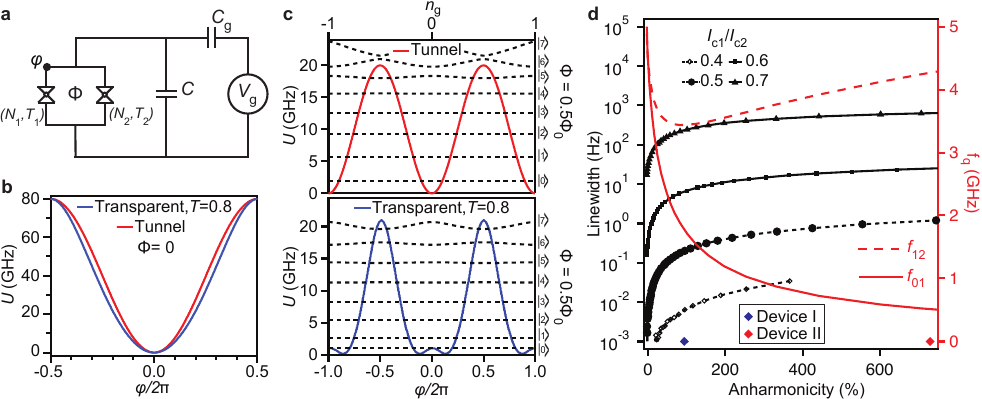}
	\caption{ \textbf{Flux-tunable gatemon simulation.} \textbf{a} Schematics of the flux-tunable gatemon qubit comprising of two superconductor-semiconductor JJs in a SQUID loop, characterized by the number of channels $N$, channel transparency $T$, and shunted by a large capacitance $C$. \textbf{b-c} Josephson potential $U$ versus the JJ phase $\varphi$ at magnetic flux $\Phi=0$ (\textbf{b}) and $\Phi = 0.5\,\Phi_0$ (\textbf{c}) for SQUID asymmetry $I_{\textrm{c1}}/I_{\textrm{c2}}=0.6$ . At $\Phi = 0.5\,\Phi_0$, the potential has a double-well shape due to the interference of Josephson energy harmonics in the SQUID. The black dashed lines represent the lowest-lying qubit energy states as a function of the offset charge $n_{\textrm{g}}$, modeled by a capacitance $C_{\textrm{g}}$ coupled to an offset voltage $V_{\textrm{g}}$ in (\textbf{a}). \textbf{d} Charge dispersion linewidth for different SQUID asymmetries $I_{\textrm{c1}}/I_{\textrm{c2}}$ (black curves) and the qubit transition frequencies for asymmetry  $I_{\textrm{c1}}/I_{\textrm{c2}}=0.7$ (red curves) as a function of the qubit anharmonicity at $\Phi = 0.5\,\Phi_{0}$. We experimentally achieve devices with anharmonicities of 96$\%$ and 730$\%$ with charge dispersion linewidths suppressed below 1 Hz, as marked by the blue and red solid circles respectively.}
	\label{fig:Fig_1}
\end{figure*}

\textit{System.} We investigate the anharmonic flux-tunable transmon qubits on an InAs quantum well (QW) proximitized by epitaxial Aluminum (epi-Al) \cite{mayer_superconducting_2019, mayer_gate_2020} at the $\Phi = 0.5 \Phi_0$ flux sweet spot. For simplicity we model the weak link JJs with their effective number of channel N and a characteristic channel transparency $T$. We will discuss more about the models of the JJ conduction in figure~\ref{fig:Fig_3}. The SQUID loop formed by two weak link JJs is shunted by a capacitance C in between the superconducting island and the ground. The qubit potential is tuned by the applied magnetic flux $\Phi$ in the SQUID loop \cite{jaklevic_quantum_1964}. We denote the phase across the first JJ as $\varphi$ and the phase of the second JJ is $2\pi \Phi/\Phi_{0} - \varphi$. As seen in figure~\ref{fig:Fig_1}a, the superconducting island is subject to voltage fluctuations modeled by a capacitively coupled electrode with an offset voltage $V_{\textrm{g}}$, which is equivalent to an offset charge of $n_{\textrm{g}} = C_{\textrm{g}}V_{\textrm{g}}/2e$ \cite{nakamura_coherent_1999}. The Hamiltonian of the qubit is given by the sum of the charging energy and the Josephson energy $U(\varphi)$:
\begin{equation}
    \label{equ:hamiltonian}
    H = 4 E_{\textrm{C}} (n-n_{\textrm{g}})^2 + U_J(\varphi),
\end{equation}
where $n$ and $\varphi$ are the number of the Cooper pairs and the phase operators, satisfying the commutator $[n,\varphi] = i$.  The Josephson energy has a simple structure for the tunnel junctions, $U_J(\varphi) = E_J(1-\cos\varphi)$. In a transmon with conventional AlOx JJs, the same scale determines the barrier height, the curvature and quartic terms at the potential minimum, resulting in competition between transmon sensitivity to charge noise and anharmonicity.

The current transport in a JJ with a semiconducting weak link is enabled by Andreev Bounds States (ABSs) formed by conduction channels in the weak link \cite{beenakker_universal_1991}. When the JJ has a superconducting phase bias $\varphi$, the energy of an ABS is given by 
\begin{equation}
    \label{equ:potentialABS}
U(\varphi, T)=\Delta \sqrt{1-T \sin^2\frac{\varphi}{2}},    
\end{equation} 
where $\Delta$ is the superconducting gap and $T$ is the conduction channel transparency. Owning to a finite lateral width in our JJs, multiple transverse channels contribute to the supercurrent across each JJ based on the lateral confinement. As such, the SQUID energy of a flux-tunable gatemon is determined by the sum over the contribution from all the ABSs in both the JJs:
\begin{equation}
    \label{equ:potentialenergy}
U_{\rm SQUID}(\varphi) =
\sum_{k=1}^{M} U(\varphi, T^{(k)}_1)+\sum_{l=1}^{N} U(2\pi\Phi/\Phi_0 - \varphi, T^{(l)}_2),
\end{equation}
where {$T^{(1)}_1$, ..., $T^{(M)}_{1}$} are the channel transparencies in the first JJ with $M$ total conduction channels and {$T^{(1)}_2$, ..., $T^{(N)}_{2}$} are the channel transparencies in the second JJ with $N$ total conduction channels. This model yields a good fit to the spectra of flux-tunable gatemon qubits with InAs nanowire JJs when the number of channels in the JJs is small \cite{hart_current-phase_2019, nichele_relating_2020, zhang_large_2024}. When the number of conduction channel is large, Eq.~\eqref{equ:potentialenergy} can be approximated by grouping the channel transparencies into one or a few characteristic transparencies or by determining the leading Fourier harmonics of the Josephson potential energies. In this study, we also compare the approximation methods mentioned above for modeling the anharmonic tunable transmon qubits with InAs 2DEG weak link JJs by examining their fits to the qubit spectra.

\begin{figure*}[]
	\centering
	\includegraphics[width=\textwidth]{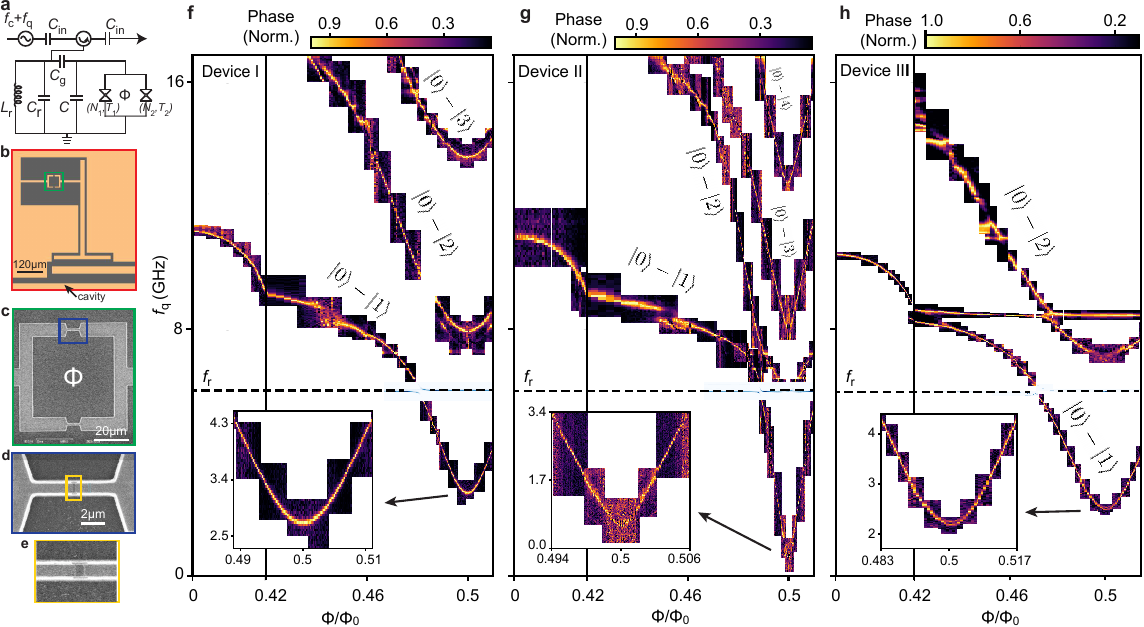}
	\caption{ \textbf{Energy spectroscopy.} \textbf{a} Circuit diagram of the flux-tunable gatemon, capacitively coupled to a $\lambda/4$ resonator and embedded in a reflection measurement setup. \textbf{b-e} False-colored scanning electron microscopy images of the qubit, with zoomed-in images of the SQUID loop and the JJ. \textbf{f-h} Normalized phase response as a function of the qubit drive frequency $f_{\textrm{q}}$ and the applied external magnetic flux $\Phi/\Phi_{0}$ for device I (\textbf{f}), device II (\textbf{g}) and device III (\textbf{h}), respectively. The black dashed line denotes the resonator frequency $f_{\textrm{r}}$ as a function of $\Phi$. \textbf{Inset:} Zoomed-in plot of the $\ket{0}-\ket{1}$ qubit transition at $\Phi = 0.5\,\Phi_{0}$, highlighting the flux 'sweet spot'.}
    \label{fig:Fig_2}
\end{figure*}

The flux-tunable gatemon qubit biased at $\Phi=0$ is equivalent to the typical gatemon qubit with similar qubit potential to a transmon qubit, as shown in figure~\ref{fig:Fig_1}b. When biased at $\Phi = 0.5 \Phi_0$, the bottom of the potential changes to a flattened double-well shape near the minima of the well, drastically increasing the qubit anharmonicity. The double-well shape of the qubit potential arises from the destructive interference of the $\cos\varphi$ terms and constructive interference of the $\cos(2\varphi)$ terms in the two JJs' potentials. The qubit states see a high potential barrier that suppresses the tunneling probability to adjacent wells. Meanwhile, the flux-tunable gatemon potential has reduced curvature of the potential near its minima so the qubit has a large anharmonicity up to a few hundred percent. As seen in figure~\ref{fig:Fig_1}c, the simulated flux-tunable gatemon $|0\rangle$ and $|1\rangle$ states are insensitive to the offset charge, but the qubit is strongly anharmonic. 

Although the trade-off between the charge dispersion and the anharmonicity still applies to the anharmonic tunable transmon qubit, the charge dispersion can be further suppressed by engineering the asymmetry of the SQUID. As shown in figure~\ref{fig:Fig_1}d, as the qubit anharmonicity increases, the charge dispersion linewidth increases by one to two orders of magnitude from a conventional transmon with less than 5\% anharmonicity. The loss in the suppression of charge dispersion can be compensated with a more asymmetric design of the SQUID, which is equivalent to increasing the first Josephson harmonic in the SQUID current-phase relation (CPR). The charge dispersion sees four orders of magnitude suppression going from $I_{\textrm{c1}}/I_{\textrm{c2}}=0.7$ to $I_{\textrm{c1}}/I_{\textrm{c2}}=0.4$, while the anharmonicity $>100\%$ remains achievable. The ability to simultaneously achieve suppressed charge dispersion and large anharmonicity for flux-tunable gatemon devices at the $\Phi = 0.5 \Phi_0$ flux sweet spot promises better qubit performance than the conventional transmon counterpart.

An equivalent circuit diagram and false color scanning electron microscopy images of the investigated anharmonic tunable transmon device are shown in figure~\ref{fig:Fig_2}a-e. We measure three nominally identical devices, all of which show similar spectra. Our typical device consists of a $\lambda/4$ resonator capacitively coupled to a common transmission line for microwave control and readout. A T-shaped Al island etched into the surrounding ground plane on an InP substrate provides the qubit shunt capacitance, which has an estimated charging energy $E_{\textrm{C}}\sim180\,$MHz from electrostatic simulations. The qubit, comprising of two weak link JJs in a SQUID loop, is connected to the T-shaped island and the ground plane via Al leads. Each individual JJ is formed by etching away a $L\sim250\,$nm long and $W\sim800$nm wide segment of the epi-Al. The qubit is read out using the $\lambda/4$ resonator with a resonance frequency of $f_{\textrm{r}}=6.01\,$GHz and Q-factor $\sim400$, and measured in a dilution refrigerator at $<50\,$mK using standard complex reflectance measurements, as shown in detail schematically in supplementary information 1. 

\textit{Spectroscopy.} We begin by measuring the normalized phase response of the readout resonator as a function of the magnetic flux $\Phi$, while sweeping the resonator drive frequency $f_{\textrm{r}}$. The resonator response for each device, indicated by the black dashed lines in Fig.~\ref{fig:Fig_2}f-h and shown in detail in supplementary information 2, exhibits a vacuum Rabi splitting with a cavity-qubit coupling strength of $g = 122\,$MHz for device I, $g = 101\,$MHz for device II and $g = 170\,$MHz for device III, as shown in supplementary information 8. These values scale with the respective charge matrix elements and are indicative of qubit states hybridizing with the readout resonator mode. To further probe the qubit transitions directly, we perform two-tone spectroscopy, where the readout resonator drive tone is fixed at its resonant frequency for each flux, while a second drive tone $f_{\textrm{q}}$ is applied to excite the qubit transitions. When biased at $\Phi=0$, each device in figure~\ref{fig:Fig_2}f-h shows a single qubit transition with transmon-like behavior, characterized by a decreasing $\ket{0}-\ket{1}$ transition frequency $f_{01}$ as $\Phi$ increases. 
In contrast, at $\Phi=0.5\,\Phi_{0}$, we observe a rich spectrum featuring multiple flux-dependent transitions that strongly anticross symmetrically around $0.5\,\Phi_{0}$. 
The qubit anharmonicity shows a strong increase with increasing flux, changing from traditional gatemon and transmon-like values of $\eta \simeq2\%$ at $\Phi=0$ to $\eta \simeq96\%$ for device I, $\eta \simeq730\%$ for device II, and $\eta \simeq125\%$ for device III at $\Phi=0.5\,\Phi_{0}$ (see Table~\ref{table:fit}). In addition, the $\ket{0}-\ket{1}$ transition shows a first-order insensitivity to flux noise at $\Phi=0.5\,\Phi_{0}$ (details in supplementary information 3), consistent with a flux `sweet spot' at half-flux quanta.  

\begin{figure*}[]
	\centering
	\includegraphics[width=\textwidth]{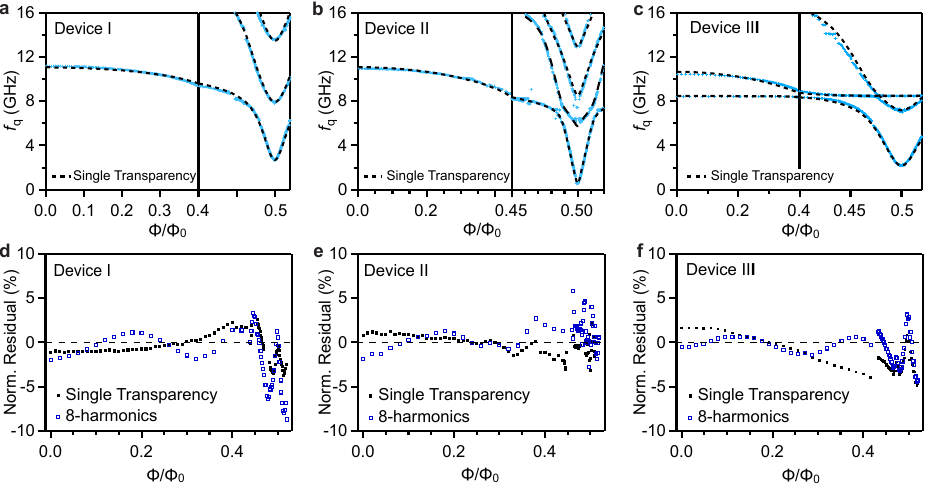}
	\caption{ \textbf{Single transparency model fit and residual.} \textbf{a-c} Qubit transition spectrum as a function of the qubit frequency $f_{\textrm{q}}$ and the applied external magnetic flux $\Phi/\Phi_{0}$ for device I (\textbf{a}), device II (\textbf{b}) and device III (\textbf{c}) extracted from figure~\ref{fig:Fig_2}f, \ref{fig:Fig_2}g and \ref{fig:Fig_2}h, respectively. The dashed black lines show fits to the qubit transitions obtained from the Hamiltonian 
    in Eq.~\eqref{equ:hamiltonian} using the single transparency model. \textbf{d-f} Normalized residual $(f_{\textrm{model}} - f_{\textrm{meas}})/f_{\textrm{meas}}$ as a function of $\Phi/\Phi_{0}$ for device I (\textbf{d}), device II (\textbf{e}) and device III (\textbf{f}), respectively.}
	\label{fig:Fig_3}
\end{figure*}

In order to understand the observed spectrum, we simulate the measured qubit transitions by determining the eigenenergies of the qubit Hamiltonian in Eq.~\eqref{equ:hamiltonian}. The qubit transition frequencies extracted from Fig.~\ref{fig:Fig_2}f-h are shown in Figs.~\ref{fig:Fig_3}a-c with blue solid curves. To account for the large number of transparent conduction channels in superconducting-semiconducting JJs, we model the Josephson potential by a single characteristic channel transparency for each JJ. This approach is justified for atomically clean JJs, where channel-to-channel variations are expected to be minimal. However, in practice, only a fraction of the total channels ($\sim10\%$ in our devices) is highly transmissive due to mesoscopic imperfections such as disorder and selective mode coupling\cite{Costas_2023, mayer_gate_2020, volkov_proximity_1993, goffman_conduction_2017}. Under this assumption, the Josephson energy
$U_{\rm SQUID}(\varphi)$ in Eq.~\ref{equ:potentialenergy} simplifies to 
$U_{\rm SQUID}(\varphi) = N_{1}U(\varphi, T_1)+N_{2}U(2\pi\Phi/\Phi_0-\varphi, T_2)$,
where $T_{1}$ and $T_{2}$ are the channel transparencies of the first and second JJ with $N_{1}$ and $N_{2}$ conduction channels, respectively. The black dashed curves in Figs.~\ref{fig:Fig_3}a-c show the calculated qubit transition frequencies with the fit parameters presented in table~\ref{table:fit}, consistent with
\begin{table}[H]
\centering
		\begin{tabular}{c||c|c|c|c|c|c|c}
			   & $N_{1} \Delta$ (GHz) & $T_{1}$ & $N_{2} \Delta$ (GHz) & $T_{2}$ & $E_{\textrm{C}}$ (GHz) & $\eta$ \\
			\hline
			I    & 221.02 $\pm$ 22.9 & 0.999 $\pm$ 0.099 & 98.90 $\pm$ 2.9 & 0.926 $\pm$ 0.006 & 0.197 $\pm$ 0.003 & 0.96\\
			II 	& 138.03 $\pm$ 3.0 & 0.928 $\pm$ 0.006 & 466.98 $\pm$ 88.5 & 0.542 $\pm$ 0.095 & 0.163 $\pm$ 0.002 & 7.3\\
			III  & 52.03 & 0.968 & 806.25 & 0.147 & 0.350 & 1.25\\
		\end{tabular}
		\caption{\textbf{Fit parameters:} channel transparency $T$, number of conduction channels $N$ and charging energy $E_{\textrm{C}}$, along with the corresponding 95$\%$ confidence intervals, determined from the single transparency model using Eq.~\eqref{equ:hamiltonian} and \ref{equ:potentialenergy}. The anharmonicity $\eta$ is extracted from measurements in figure~\ref{fig:Fig_2}f-h. The details of the error analysis are described in supplementary information 9.} 
	\label{table:fit}
\end{table} 
the short-junction limit and quasi-ballistic regime estimated in supplementary information 11 for our devices. In addition, our model indicates a strong suppression of the charge matrix element, as shown in supplementary information 4, for the $\ket{0}-\ket{1}$ qubit transition at $0.5\,\Phi_{0}$. This suppression diminishes the dipole coupling of the qubit to charge noise and dielectric loss channels, thereby enhancing the $T_{1}$ relaxation time at the half-flux sweet spot.

To quantitatively assess the accuracy of the single transparency model, we plot the normalized residual $(f_{\textrm{model}} - f_{\textrm{meas}})/f_{\textrm{meas}}$ vs $\Phi$ for the lowest-energy $\ket{0}-\ket{1}$ transition in Fig.~\ref{fig:Fig_3}d-f. As shown by the black filled squares, we observe a residual of $<2\%$ at $\Phi=0$ and no more than $4\%$ at $\Phi = 0.5\,\Phi_{0}$ for device I and II, while device III remains below $5\%$ for all $\Phi$. For comparison, we consider a higher-harmonic model in which the Josephson potential energy for each JJ is approximated by its leading Fourier harmonic contributions: $U_{i} = \sum_{k} E_{\textrm{J}i}^{k} \cos{(k\varphi)}$, where $i\in{[1,2]}$ refers to the first and second JJ. The normalized residual, determined using the four leading harmonic terms $k=4$ (see supplementary information 5 for $k=1$ and $k=2$) for each JJ, exhibits an oscillating behavior with residual $\leq 10\%$ for device I and $\leq 5\%$ for devices II and III (blue hollow squares). These results indicate that the single transparency model provides the most accurate representation of the flux-tunable gatemon spectra despite using fewer fitting parameters.  We observe that further increasing the number of characteristic transparencies does not significantly reduce the residual, as demonstrated in supplementary information 6.    

\begin{figure*}[h]
	\centering
	\includegraphics[width=\textwidth]{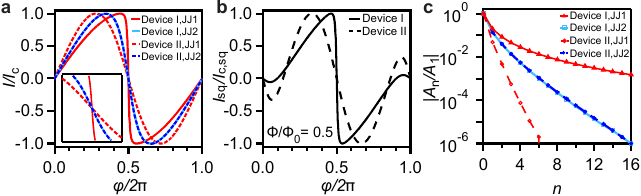}
	\caption{ \textbf{Current phase relation (CPR).} \textbf{a} Current phase relation ($I/I_{c}$ vs $\varphi$) of each JJ in device I and II, determined using the fit parameters obtained from the single transparency model in table~\ref{table:fit}. We denote the JJ with larger critical current as JJ1 in each device. \textbf{Inset:} Zoomed-in image of the CPRs at $\varphi/2\pi = 0.5$. \textbf{b} CPR of the SQUID loop ($I_{\textrm{sq}}/I_{\textrm{c,sq}}$ vs $\varphi$) at $\Phi = 0.5\,\Phi_{0}$ calculated using Eq.~\eqref{equ:squidIc}. \textbf{c} Normalized $n^{\textrm{th}}$-order harmonic contribution $|{A_{\textrm{n}}/A_{1}}|$ for the leading 16 terms, i.e. up to $n=16$, with an amplitude baseline threshold set at $|{A_{\textrm{n}}/A_{1}}| = 10^{-6}$.}
	\label{fig:Fig_4}
\end{figure*}

The extracted fit parameters allows us to determine the CPR of the individual JJs and the SQUID loop. For each JJ, we use the Josephson potential: $I_{i}(\varphi_{i}) = \frac{2\pi}{\Phi_{0}} \frac{\partial U_{i}}{\partial\varphi_{i}}$ with $i \in [{1,2}]$ for JJ1 and JJ2 respectively, to extract the normalized CPR $I/I_{\textrm{c}}$ for devices I and II in figure~\ref{fig:Fig_4}a. We observe highly skewed non-sinusoidal CPRs, reminiscent of higher-order harmonic contributions due to the interference of supercurrents carried by 2-electrons, 4-electrons and higher charges across the JJs. Consequently, for the SQUID loop CPR at $\Phi = 0.5\,\Phi_{0}$, we sum over the individual JJ contributions to obtain:
\begin{equation}
    \label{equ:squidIc}
    \begin{split}
        I_{\textrm{sq}} (\varphi) = I_{1} (\varphi) + I_{2}(2\pi\frac{\Phi}{\Phi_{0}} - \varphi) 
    \end{split}
\end{equation}
$I_{\textrm{sq}}/I_{\textrm{c,sq}}$ extracted for devices I and II is shown in Fig.~\ref{fig:Fig_4}b. We further extract the higher-order harmonic contributions for each JJ's CPR using: $I_{\textrm{i}} (\varphi_{i}) = \sum_{n} A_{n}\sin{(n\varphi)}$. Figure~\ref{fig:Fig_4}c shows the normalized $n^{\textrm{th}}$-order harmonic term $|A_{\textrm{n}}/A_{1}|$ for devices I and II, determined up to the leading sixteen terms ($n=16$) with an amplitude baseline threshold set at $|A_{\textrm{n}}/A_{1}| = 10^{-6}$. We observe a gradual fall-off in the harmonic amplitudes with increasing $n$. In addition, while some JJs exhibit similar CPRs, others show substantial deviations, resulting in distinct harmonic content. This diversity reflects the sensitivity of the CPR to microscopic transport parameters of the JJs. The harmonic analysis provides complementary information to standard transport measurements, allowing one to identify deviations from purely sinusoidal behavior and engineer the Josephson potential landscape including the realization of multi-well potentials, thereby tailoring qubit properties for novel circuit architectures and optimized performance. 



\begin{figure}[]
	\centering
	\includegraphics[width=0.85\textwidth]{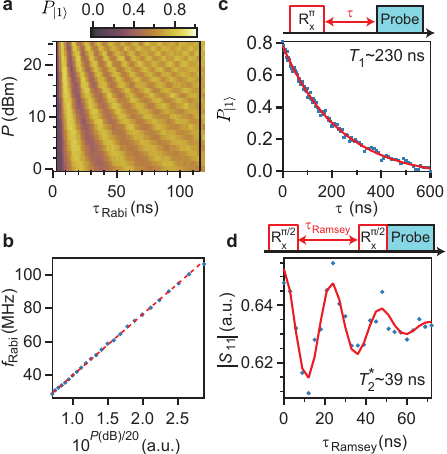}
	\caption{\textbf{Coherent qubit control and manipulation.} \textbf{a} Rabi oscillations as a function of the qubit drive power $P$ and the Rabi pulse duration $\tau_{\textrm{Rabi}}$ at $0.5\,\Phi_{0}$. \textbf{b} Rabi frequency $f_{\textrm{Rabi}}$ extracted from (\textbf{a}) as a function of $10^{P(\textrm{dB})/20}$, which is proportional to the qubit drive amplitude $\sqrt{P}$. \textbf{c} Energy relaxation time $T_{1}$ measured at $\Phi = 0.5\,\Phi_{0}$. The solid red curve is an exponential fit with time constant $T_{1} \sim 230\,$ns. \textbf{d} Ramsey coherence time $T_{2}^{*} \sim39\,$ns measured at $\Phi = 0.5\,\Phi_{0}$.}
	\label{fig:Fig_5}
\end{figure}

\textit{Qubit coherence properties.} We now demonstrate the basic operations of the flux-tunable gatemon qubit by performing Rabi oscillation measurements and Ramsey interferometry using time-domain manipulation and readout. For each measurement, microwave pulses with frequency $f_{\textrm{q}} \approx f_{01}$ are applied via the cavity readout feedline, after an initialization time of $100\,\mu$s to allow the qubit to relax to the ground state $\ket{0}$. For Rabi measurements, we apply the drive pulse for a time duration $\tau_{\textrm{Rabi}}$, followed by reading out the qubit state via the cavity. The microwave pulse induces Rabi oscillations between the qubit states $\ket{0}$ and $\ket{1}$, as shown in figure~\ref{fig:Fig_5}a, where we plot the probability of qubit state $\ket{1}$, $P_{\ket{1}}$ as a function of $\tau_{\textrm{Rabi}}$ and the applied drive power $P$. We observe that the Rabi frequency increases for larger drive power $P$. To explicitly demonstrate this, we extract the Rabi frequency $f_{\textrm{Rabi}}$ for each power $P$ by fitting the Rabi oscillations in figure~\ref{fig:Fig_5}a with a sinusoidal function with exponentially decaying envelope: $A\sin{(2\pi f_{\textrm{Rabi}}t)}e^{-t/\tau_{\textrm{Rabi}}}$. The extracted $f_{\textrm{Rabi}}$ as a function of $10^{P(\textrm{dBm})/20}$, which is proportional to the drive amplitude $\sqrt{P}$, shows a linear dependence in figure~\ref{fig:Fig_5}b, with Rabi frequencies as large as $>100\,$MHz enabling coherent manipulation and gate operations of the flux-tunable gatemon qubit much faster than traditional transmons and gatemons.

We first measure the energy relaxation time $T_{1}$ to estimate the qubit coherence times. The Rabi oscillations data in figure~\ref{fig:Fig_5}a allow us to calibrate the pulse amplitudes and durations for the corresponding rotations around the x-axis on the Bloch sphere. To measure $T_{1}$, we first apply a $R^{\pi}_{\textrm{X}}$ pulse to excite the qubit to state $\ket{1}$, followed by a waiting time delay $\tau$ before readout. The qubit probability $P_{\ket{1}}$, plotted in figure~\ref{fig:Fig_5}c, as a function of the time delay $\tau$ shows an exponential decay due to the qubit relaxation, yielding $T_{1}\sim230\,$ns for device I. Similarly, to measure Ramsey dephasing time $T_{2}^{*}$, two $R^{\pi/2}_{\textrm{X}}$ pulses slightly detuned from the qubit frequency, $\delta f\sim40\,$MHz and separated by a time delay $\tau_{\textrm{Ramsey}}$ are applied before readout. The resulting cavity readout response as a function of the time delay $\tau_{\textrm{Ramsey}}$, shown in figure~\ref{fig:Fig_5}d, demonstrates Ramsey fringes, consistent with the qubit state acquiring a phase $\varphi = 2\pi\delta {f}\tau_{\textrm{Ramsey}}$ while precessing around the $z$-axis of the Bloch sphere. We obtain $T^{*}_{2}\sim39\,$ns by fitting to a sinusoidal function with an exponential decay envelope. 

As a last step, we critically evaluate the dominant loss mechanisms affecting the qubit's relaxation and decoherence times. We observe that $T_{1}$ is independent of the qubit frequency $f_{01}$ in the vicinity of $\Phi = 0.5\,\Phi$ (shown in supplementary figure 7), which we then use to determine the effective dielectric loss tangent $\tan\delta_{\textrm{c}}$ using the charge matrix elements from supplementary information 4. We obtain $\tan\delta_{\textrm{c}}$ to be on the order of $\sim10^{-4}$, consistent with typical values for dielectrics such as InP and AlOx, but significantly higher than the typical values of $\sim10^{-6}$ observed for traditional transmon devices on silicon and sapphire substrates. This suggests that dielectric losses, likely originating from the InP host substrate, limit $T_{1}$. Additionally, we note that the dephasing time $T_{2} \ll 2T_{1}$, indicating that qubit coherence is not primarily limited by energy relaxation, but rather by other on-chip dephasing mechanisms, for example, by 1/$f$ noise from critical current fluctuations\cite{koch_charge-insensitive_2007} or the decoherence of the Andreev Bound States close to the $\varphi=\pi$ phase bias\cite{janvier_coherent_2015}.  

In conclusion, we demonstrate a hybrid InAs-Al flux-tunable transmon qubit biased at the half-integer flux sweet spot, which exhibits a significant enhancement in anharmonicity - over an order of magnitude increase - without any electrical gating, along with a minimal charge dispersion, thereby reducing its sensitivity to offset charge noise. Additionally in our model, we observe a suppressed charge matrix element by a factor of 2, enabling partial parity protection of the qubit without the need for strict symmetry between the two JJs'. This eliminates the need for precise fabrication with accurate electrical gating or additional circuit elements\cite{banszerus_hybrid_2024, leblanc_gate-_2025} necessary to maintain the symmetry. Such large intrinsic qubit anharmonicities enables fast single-gate times of less than 10 ns, surpassing the Rabi frequencies typically achieved in traditional gatemons and transmons without any pulse shaping. Although dielectric losses currently limit the qubit lifetime, the use of less lossy substrate materials, for example, Si or sapphire, should make it possible to achieve much longer coherence and energy relaxation times. Our work is versatile enough to be implemented in other material systems such as Ge or SiGe, and provides a novel approach to characterize various superconducting-semiconducting material platforms and implement strongly anharmonic, partially protected qubits for quantum information processing without the need for complex design and fabrication.

\section*{METHODS}
The anharmonic tunable transmon qubit device is fabricated on a stacked heterostructure grown on $500\,\mu$m thick Fe-doped InP substrate. An In$_{0.53}$Al$_{0.47}$/In$_{0.52}$Al$_{0.48}$ superlattice and a graded In$_x$Al$_{1-x}$As layer are further grown on the substrate to act as a buffer layer, followed by 4nm of InGaAs, 4nm of InAs and 10nm of InGaAs, which defines the quantum well structure with InAs as the 2DEG layer. The wafer is then capped with a $25-30\,$nm thick layer of Al deposited $in-situ$ with molecular beam epitaxy. 

The coplanar waveguide (CPW) resonators are fabricated on a $6\,$mm by $5\,$mm rectangular wafer with a center conductor width of 35 $\mu$m and a trench gap width of 20$\mu$m using e-beam lithography. The capping Al layer is then removed with 50$^{\circ}$C Transene Aluminum Etchant Type D, followed by a wet-etch of the III-V layers at room-temperature for 3-4 minutes, resulting in a trench depth of $350-450\,$nm. The III-V layer etchant is a mixture of phosphoric acid (85\%), hydrogen peroxide (30\%) and DI water with a volumetric ratio of 1:2:60 respectively. The qubit SQUID, fabricated in the same step as the CPW resonators, has a dimension of 50$\mu$m by 50$\mu$m for devices 1 and 2, and 22$\mu$m by 22$\mu$m for device 3. We then etch the Al for 6 seconds in a separate e-beam lithography step to form the Josephson junctions. The device is finally packaged in a copper cavity via Al bond wires and measured in a dilution refrigeration at $<50\,$mK using standard reflectance measurements.

\section*{Data Availability Statement}
\noindent All data in this publication are available in numerical form at \url{https://doi.org/10.5281/zenodo.17654710}.

\bibliography{references}

@article{janvier_coherent_2015,
	title = {Coherent manipulation of {Andreev} states in superconducting atomic contacts},
	volume = {349},
	url = {https://www.science.org/doi/10.1126/science.aab2179},
	doi = {10.1126/science.aab2179},
	number = {6253},
	urldate = {2025-02-25},
	journal = {Science},
	author = {Janvier, C. and Tosi, L. and Bretheau, L. and Girit, Ç. Ö. and Stern, M. and Bertet, P. and Joyez, P. and Vion, D. and Esteve, D. and Goffman, M. F. and Pothier, H. and Urbina, C.},
	month = sep,
	year = {2015},
	note = {Publisher: American Association for the Advancement of Science},
	pages = {1199--1202},
}

@article{houck_life_2009,
	title = {Life after charge noise: recent results with transmon qubits},
	volume = {8},
	issn = {1573-1332},
	shorttitle = {Life after charge noise},
	url = {https://doi.org/10.1007/s11128-009-0100-6},
	doi = {10.1007/s11128-009-0100-6},
	number = {2},
	urldate = {2025-02-23},
	journal = {Quantum Information Processing},
	author = {Houck, A. A. and Koch, Jens and Devoret, M. H. and Girvin, S. M. and Schoelkopf, R. J.},
	month = jun,
	year = {2009},
	pages = {105--115},
}

@article{schreier_suppressing_2008,
	title = {Suppressing charge noise decoherence in superconducting charge qubits},
	volume = {77},
	url = {https://link.aps.org/doi/10.1103/PhysRevB.77.180502},
	doi = {10.1103/PhysRevB.77.180502},
	number = {18},
	urldate = {2025-02-23},
	journal = {Physical Review B},
	author = {Schreier, J. A. and Houck, A. A. and Koch, Jens and Schuster, D. I. and Johnson, B. R. and Chow, J. M. and Gambetta, J. M. and Majer, J. and Frunzio, L. and Devoret, M. H. and Girvin, S. M. and Schoelkopf, R. J.},
	month = may,
	year = {2008},
	note = {Publisher: American Physical Society},
	pages = {180502},
}

@article{jaklevic_quantum_1964,
	title = {Quantum {Interference} {Effects} in {Josephson} {Tunneling}},
	volume = {12},
	url = {https://link.aps.org/doi/10.1103/PhysRevLett.12.159},
	doi = {10.1103/PhysRevLett.12.159},
	number = {7},
	urldate = {2025-02-23},
	journal = {Physical Review Letters},
	author = {Jaklevic, R. C. and Lambe, John and Silver, A. H. and Mercereau, J. E.},
	month = feb,
	year = {1964},
	note = {Publisher: American Physical Society},
	pages = {159--160},
}

@article{houck_controlling_2008,
	title = {Controlling the {Spontaneous} {Emission} of a {Superconducting} {Transmon} {Qubit}},
	volume = {101},
	url = {https://link.aps.org/doi/10.1103/PhysRevLett.101.080502},
	doi = {10.1103/PhysRevLett.101.080502},
	number = {8},
	urldate = {2025-02-23},
	journal = {Physical Review Letters},
	author = {Houck, A. A. and Schreier, J. A. and Johnson, B. R. and Chow, J. M. and Koch, Jens and Gambetta, J. M. and Schuster, D. I. and Frunzio, L. and Devoret, M. H. and Girvin, S. M. and Schoelkopf, R. J.},
	month = aug,
	year = {2008},
	note = {Publisher: American Physical Society},
	pages = {080502},
}

@article{chow_randomized_2009,
	title = {Randomized {Benchmarking} and {Process} {Tomography} for {Gate} {Errors} in a {Solid}-{State} {Qubit}},
	volume = {102},
	url = {https://link.aps.org/doi/10.1103/PhysRevLett.102.090502},
	doi = {10.1103/PhysRevLett.102.090502},
	number = {9},
	urldate = {2025-02-23},
	journal = {Physical Review Letters},
	author = {Chow, J. M. and Gambetta, J. M. and Tornberg, L. and Koch, Jens and Bishop, Lev S. and Houck, A. A. and Johnson, B. R. and Frunzio, L. and Girvin, S. M. and Schoelkopf, R. J.},
	month = mar,
	year = {2009},
	note = {Publisher: American Physical Society},
	pages = {090502},
}

@article{steffen_accurate_2003,
	title = {Accurate control of {Josephson} phase qubits},
	volume = {68},
	url = {https://link.aps.org/doi/10.1103/PhysRevB.68.224518},
	doi = {10.1103/PhysRevB.68.224518},
	number = {22},
	urldate = {2025-02-23},
	journal = {Physical Review B},
	author = {Steffen, Matthias and Martinis, John M. and Chuang, Isaac L.},
	month = dec,
	year = {2003},
	note = {Publisher: American Physical Society},
	pages = {224518},
}

@article{pita-vidal_gate-tunable_2020,
	title = {Gate-{Tunable} {Field}-{Compatible} {Fluxonium}},
	volume = {14},
	url = {https://link.aps.org/doi/10.1103/PhysRevApplied.14.064038},
	doi = {10.1103/PhysRevApplied.14.064038},
	number = {6},
	urldate = {2025-02-23},
	journal = {Physical Review Applied},
	author = {Pita-Vidal, Marta and Bargerbos, Arno and Yang, Chung-Kai and van Woerkom, David J. and Pfaff, Wolfgang and Haider, Nadia and Krogstrup, Peter and Kouwenhoven, Leo P. and de Lange, Gijs and Kou, Angela},
	month = dec,
	year = {2020},
	note = {Publisher: American Physical Society},
	pages = {064038},
}

@article{huo_gatemon_2023,
	title = {Gatemon {Qubit} {Based} on a {Thin} {InAs}-{Al} {Hybrid} {Nanowire}},
	volume = {40},
	issn = {0256-307X},
	url = {https://dx.doi.org/10.1088/0256-307X/40/4/047302},
	doi = {10.1088/0256-307X/40/4/047302},
	number = {4},
	urldate = {2025-02-23},
	journal = {Chinese Physics Letters},
	author = {Huo, Jierong and Xia, Zezhou and Li, Zonglin and Zhang, Shan and Wang, Yuqing and Pan, Dong and Liu, Qichun and Liu, Yulong and Wang, Zhichuan and Gao, Yichun and Zhao, Jianhua and Li, Tiefu and Ying, Jianghua and Shang, Runan and Zhang, Hao},
	month = mar,
	year = {2023},
	note = {Publisher: Chinese Physical Society and IOP Publishing Ltd},
	pages = {047302},
}

@article{kiyooka_gatemon_2025,
	title = {Gatemon {Qubit} on a {Germanium} {Quantum}-{Well} {Heterostructure}},
	volume = {25},
	issn = {1530-6984},
	url = {https://doi.org/10.1021/acs.nanolett.4c05539},
	doi = {10.1021/acs.nanolett.4c05539},
	number = {1},
	urldate = {2025-02-23},
	journal = {Nano Letters},
	author = {Kiyooka, Elyjah and Tangchingchai, Chotivut and Noirot, Leo and Leblanc, Axel and Brun, Boris and Zihlmann, Simon and Maurand, Romain and Schmitt, Vivien and Dumur, Étienne and Hartmann, Jean-Michel and Lefloch, Francois and De Franceschi, Silvano},
	month = jan,
	year = {2025},
	note = {Publisher: American Chemical Society},
	pages = {562--568},
}

@article{casparis_gatemon_2016,
	title = {Gatemon {Benchmarking} and {Two}-{Qubit} {Operations}},
	volume = {116},
	url = {https://link.aps.org/doi/10.1103/PhysRevLett.116.150505},
	doi = {10.1103/PhysRevLett.116.150505},
	number = {15},
	urldate = {2025-02-23},
	journal = {Physical Review Letters},
	author = {Casparis, L. and Larsen, T. W. and Olsen, M. S. and Kuemmeth, F. and Krogstrup, P. and Nygård, J. and Petersson, K. D. and Marcus, C. M.},
	month = apr,
	year = {2016},
	note = {Publisher: American Physical Society},
	pages = {150505},
}

@article{fominov_asymmetric_2022,
	title = {Asymmetric higher-harmonic {SQUID} as a {Josephson} diode},
	volume = {106},
	url = {https://link.aps.org/doi/10.1103/PhysRevB.106.134514},
	doi = {10.1103/PhysRevB.106.134514},
	number = {13},
	urldate = {2025-02-12},
	journal = {Physical Review B},
	author = {Fominov, Ya. V. and Mikhailov, D. S.},
	month = oct,
	year = {2022},
	note = {Publisher: American Physical Society},
	pages = {134514},
}

@article{beenakker_universal_1991,
	title = {Universal limit of critical-current fluctuations in mesoscopic {Josephson} junctions},
	volume = {67},
	url = {https://link.aps.org/doi/10.1103/PhysRevLett.67.3836},
	doi = {10.1103/PhysRevLett.67.3836},
	number = {27},
	urldate = {2025-02-12},
	journal = {Physical Review Letters},
	author = {Beenakker, C. W. J.},
	month = dec,
	year = {1991},
	note = {Publisher: American Physical Society},
	pages = {3836--3839},
}

@article{leblanc_gate-_2025,
	title = {Gate- and flux-tunable sin(2φ) {Josephson} element with planar-{Ge} junctions},
	volume = {16},
	copyright = {2025 The Author(s)},
	issn = {2041-1723},
	url = {https://www.nature.com/articles/s41467-025-56245-7},
	doi = {10.1038/s41467-025-56245-7},
	number = {1},
	urldate = {2025-03-15},
	journal = {Nature Communications},
	author = {Leblanc, Axel and Tangchingchai, Chotivut and Sadre Momtaz, Zahra and Kiyooka, Elyjah and Hartmann, Jean-Michel and Gustavo, Frédéric and Thomassin, Jean-Luc and Brun, Boris and Schmitt, Vivien and Zihlmann, Simon and Maurand, Romain and Dumur, Étienne and De Franceschi, Silvano and Lefloch, François},
	month = jan,
	year = {2025},
	note = {Publisher: Nature Publishing Group},
	pages = {1010},
	}

@article{nakamura_coherent_1999,
	title = {Coherent control of macroscopic quantum states in a single-{Cooper}-pair box},
	volume = {398},
	copyright = {1999 Macmillan Magazines Ltd.},
	issn = {1476-4687},
	url = {https://www.nature.com/articles/19718},
	doi = {10.1038/19718},
	number = {6730},
	urldate = {2025-01-16},
	journal = {Nature},
	author = {Nakamura, Y. and Pashkin, Yu A. and Tsai, J. S.},
	month = apr,
	year = {1999},
	note = {Publisher: Nature Publishing Group},
	pages = {786--788},
}

@article{kringhoj_anharmonicity_2018,
	title = {Anharmonicity of a superconducting qubit with a few-mode {Josephson} junction},
	volume = {97},
	url = {https://link.aps.org/doi/10.1103/PhysRevB.97.060508},
	doi = {10.1103/PhysRevB.97.060508},
	number = {6},
	urldate = {2025-01-16},
	journal = {Physical Review B},
	author = {Kringhøj, A. and Casparis, L. and Hell, M. and Larsen, T. W. and Kuemmeth, F. and Leijnse, M. and Flensberg, K. and Krogstrup, P. and Nygård, J. and Petersson, K. D. and Marcus, C. M.},
	month = feb,
	year = {2018},
	note = {Publisher: American Physical Society},
	pages = {060508},
}

@article{luthi_evolution_2018,
	title = {Evolution of {Nanowire} {Transmon} {Qubits} and {Their} {Coherence} in a {Magnetic} {Field}},
	volume = {120},
	url = {https://link.aps.org/doi/10.1103/PhysRevLett.120.100502},
	doi = {10.1103/PhysRevLett.120.100502},
	number = {10},
	urldate = {2025-01-16},
	journal = {Physical Review Letters},
	author = {Luthi, F. and Stavenga, T. and Enzing, O. W. and Bruno, A. and Dickel, C. and Langford, N. K. and Rol, M. A. and Jespersen, T. S. and Nygård, J. and Krogstrup, P. and DiCarlo, L.},
	month = mar,
	year = {2018},
	note = {Publisher: American Physical Society},
	pages = {100502},
}

@article{hertel_gate-tunable_2022,
	title = {Gate-{Tunable} {Transmon} {Using} {Selective}-{Area}-{Grown} {Superconductor}-{Semiconductor} {Hybrid} {Structures} on {Silicon}},
	volume = {18},
	url = {https://link.aps.org/doi/10.1103/PhysRevApplied.18.034042},
	doi = {10.1103/PhysRevApplied.18.034042},
	number = {3},
	urldate = {2025-01-16},
	journal = {Physical Review Applied},
	author = {Hertel, Albert and Eichinger, Michaela and Andersen, Laurits O. and van Zanten, David M.T. and Kallatt, Sangeeth and Scarlino, Pasquale and Kringhøj, Anders and Chavez-Garcia, José M. and Gardner, Geoffrey C. and Gronin, Sergei and Manfra, Michael J. and Gyenis, András and Kjaergaard, Morten and Marcus, Charles M. and Petersson, Karl D.},
	month = sep,
	year = {2022},
	note = {Publisher: American Physical Society},
	pages = {034042},
}

@article{sagi_gate_2024,
	title = {A gate tunable transmon qubit in planar {Ge}},
	volume = {15},
	copyright = {2024 The Author(s)},
	issn = {2041-1723},
	url = {https://www.nature.com/articles/s41467-024-50763-6},
	doi = {10.1038/s41467-024-50763-6},
	language = {en},
	number = {1},
	urldate = {2025-01-16},
	journal = {Nature Communications},
	author = {Sagi, Oliver and Crippa, Alessandro and Valentini, Marco and Janik, Marian and Baghumyan, Levon and Fabris, Giorgio and Kapoor, Lucky and Hassani, Farid and Fink, Johannes and Calcaterra, Stefano and Chrastina, Daniel and Isella, Giovanni and Katsaros, Georgios},
	month = jul,
	year = {2024},
	note = {Publisher: Nature Publishing Group},
	pages = {6400},
}

@article{koch_charge-insensitive_2007,
	title = {Charge-insensitive qubit design derived from the {Cooper} pair box},
	volume = {76},
	url = {https://link.aps.org/doi/10.1103/PhysRevA.76.042319},
	doi = {10.1103/PhysRevA.76.042319},
	number = {4},
	urldate = {2025-01-16},
	journal = {Physical Review A},
	author = {Koch, Jens and Yu, Terri M. and Gambetta, Jay and Houck, A. A. and Schuster, D. I. and Majer, J. and Blais, Alexandre and Devoret, M. H. and Girvin, S. M. and Schoelkopf, R. J.},
	month = oct,
	year = {2007},
	note = {Publisher: American Physical Society},
	pages = {042319},
}

@article{mayer_superconducting_2019,
	title = {Superconducting proximity effect in epitaxial {Al}-{InAs} heterostructures},
	volume = {114},
	issn = {0003-6951},
	url = {https://doi.org/10.1063/1.5067363},
	doi = {10.1063/1.5067363},
	number = {10},
	urldate = {2025-01-07},
	journal = {Applied Physics Letters},
	author = {Mayer, William and Yuan, Joseph and Wickramasinghe, Kaushini S. and Nguyen, Tri and Dartiailh, Matthieu C. and Shabani, Javad},
	month = mar,
	year = {2019},
	pages = {103104},
}

@article{strickland_gatemonium_2024,
	title = {Gatemonium: A Voltage-Tunable Fluxonium},
	author = {Strickland, William M. and Elfeky, Bassel Heiba and Baker, Lukas and Maiani, Andrea and Lee, Jaewoo and Levy, Ido and Issokson, Jacob and Vrajitoarea, Andrei and Shabani, Javad},
	journal = {PRX Quantum},
	volume = {6},
	issue = {1},
	pages = {010326},
	numpages = {15},
	year = {2025},
	month = {Feb},
	publisher = {American Physical Society},
	doi = {10.1103/PRXQuantum.6.010326},
	url = {https://link.aps.org/doi/10.1103/PRXQuantum.6.010326}
}

@article{ciaccia_charge-4e_2024,
	title = {Charge-4e supercurrent in a two-dimensional {InAs}-{Al} superconductor-semiconductor heterostructure},
	volume = {7},
	copyright = {2024 The Author(s)},
	issn = {2399-3650},
	url = {https://www.nature.com/articles/s42005-024-01531-x},
	doi = {10.1038/s42005-024-01531-x},
	language = {en},
	number = {1},
	urldate = {2025-01-07},
	journal = {Communications Physics},
	author = {Ciaccia, Carlo and Haller, Roy and Drachmann, Asbjørn C. C. and Lindemann, Tyler and Manfra, Michael J. and Schrade, Constantin and Schönenberger, Christian},
	month = jan,
	year = {2024},
	note = {Publisher: Nature Publishing Group},
	pages = {1--8},
}

@article{zhang_large_2024,
	title = {Large {Second}-{Order} {Josephson} {Effect} in {Planar} {Superconductor}-{Semiconductor} {Junctions}},
	volume = {16},
	issn = {2542-4653},
	url = {http://arxiv.org/abs/2211.07119},
	doi = {10.21468/SciPostPhys.16.1.030},
	number = {1},
	urldate = {2025-01-07},
	journal = {SciPost Physics},
	author = {Zhang, P. and Zarassi, A. and Jarjat, L. and Sande, V. Van de and Pendharkar, M. and Lee, J. S. and Dempsey, C. P. and McFadden, A. P. and Harrington, S. D. and Dong, J. T. and Wu, H. and Chen, A.-H. and Hocevar, M. and Palmstrøm, C. J. and Frolov, S. M.},
	month = jan,
	year = {2024},
	note = {arXiv:2211.07119 [cond-mat]},
	pages = {030},
}

@article{de_lange_realization_2015,
	title = {Realization of {Microwave} {Quantum} {Circuits} {Using} {Hybrid} {Superconducting}-{Semiconducting} {Nanowire} {Josephson} {Elements}},
	volume = {115},
	url = {https://link.aps.org/doi/10.1103/PhysRevLett.115.127002},
	doi = {10.1103/PhysRevLett.115.127002},
	number = {12},
	urldate = {2025-01-06},
	journal = {Physical Review Letters},
	author = {de Lange, G. and van Heck, B. and Bruno, A. and van Woerkom, D. J. and Geresdi, A. and Plissard, S. R. and Bakkers, E. P. A. M. and Akhmerov, A. R. and DiCarlo, L.},
	month = sep,
	year = {2015},
	note = {Publisher: American Physical Society},
	pages = {127002},
}

@article{hart_current-phase_2019,
	title = {Current-phase relations of {InAs} nanowire {Josephson} junctions: {From} interacting to multimode regimes},
	volume = {100},
	shorttitle = {Current-phase relations of {InAs} nanowire {Josephson} junctions},
	url = {https://link.aps.org/doi/10.1103/PhysRevB.100.064523},
	doi = {10.1103/PhysRevB.100.064523},
	number = {6},
	urldate = {2025-01-06},
	journal = {Physical Review B},
	author = {Hart, Sean and Cui, Zheng and Ménard, Gerbold and Deng, Mingtang and Antipov, Andrey E. and Lutchyn, Roman M. and Krogstrup, Peter and Marcus, Charles M. and Moler, Kathryn A.},
	month = aug,
	year = {2019},
	note = {Publisher: American Physical Society},
	pages = {064523},
}

@article{nichele_relating_2020,
	title = {Relating {Andreev} {Bound} {States} and {Supercurrents} in {Hybrid} {Josephson} {Junctions}},
	volume = {124},
	url = {https://link.aps.org/doi/10.1103/PhysRevLett.124.226801},
	doi = {10.1103/PhysRevLett.124.226801},
	number = {22},
	urldate = {2025-01-06},
	journal = {Physical Review Letters},
	author = {Nichele, F. and Portolés, E. and Fornieri, A. and Whiticar, A. M. and Drachmann, A. C. C. and Gronin, S. and Wang, T. and Gardner, G. C. and Thomas, C. and Hatke, A. T. and Manfra, M. J. and Marcus, C. M.},
	month = jun,
	year = {2020},
	note = {Publisher: American Physical Society},
	pages = {226801},
}

@article{erlandsson_parity_2023,
	title = {Parity switching in a full-shell superconductor-semiconductor nanowire qubit},
	volume = {108},
	url = {https://link.aps.org/doi/10.1103/PhysRevB.108.L121406},
	doi = {10.1103/PhysRevB.108.L121406},
	number = {12},
	urldate = {2025-01-06},
	journal = {Physical Review B},
	author = {Erlandsson, O. and Sabonis, D. and Kringhøj, A. and Larsen, T. W. and Krogstrup, P. and Petersson, K. D. and Marcus, C. M.},
	month = sep,
	year = {2023},
	note = {Publisher: American Physical Society},
	pages = {L121406},
}

@article{kringhoj_suppressed_2020,
	title = {Suppressed {Charge} {Dispersion} via {Resonant} {Tunneling} in a {Single}-{Channel} {Transmon}},
	volume = {124},
	url = {https://link.aps.org/doi/10.1103/PhysRevLett.124.246803},
	doi = {10.1103/PhysRevLett.124.246803},
	number = {24},
	urldate = {2024-10-30},
	journal = {Physical Review Letters},
	author = {Kringhøj, A. and van Heck, B. and Larsen, T. W. and Erlandsson, O. and Sabonis, D. and Krogstrup, P. and Casparis, L. and Petersson, K. D. and Marcus, C. M.},
	month = jun,
	year = {2020},
	note = {Publisher: American Physical Society},
	pages = {246803},
}

@article{bargerbos_observation_2020,
	title = {Observation of {Vanishing} {Charge} {Dispersion} of a {Nearly} {Open} {Superconducting} {Island}},
	volume = {124},
	url = {https://link.aps.org/doi/10.1103/PhysRevLett.124.246802},
	doi = {10.1103/PhysRevLett.124.246802},
	number = {24},
	urldate = {2024-10-30},
	journal = {Physical Review Letters},
	author = {Bargerbos, Arno and Uilhoorn, Willemijn and Yang, Chung-Kai and Krogstrup, Peter and Kouwenhoven, Leo P. and de Lange, Gijs and van Heck, Bernard and Kou, Angela},
	month = jun,
	year = {2020},
	note = {Publisher: American Physical Society},
	pages = {246802},
}

@article{danilenko_few-mode_2023,
	title = {Few-mode to mesoscopic junctions in gatemon qubits},
	volume = {108},
	url = {https://link.aps.org/doi/10.1103/PhysRevB.108.L020505},
	doi = {10.1103/PhysRevB.108.L020505},
	number = {2},
	urldate = {2025-01-06},
	journal = {Physical Review B},
	author = {Danilenko, Alisa and Sabonis, Deividas and Winkler, Georg W. and Erlandsson, Oscar and Krogstrup, Peter and Marcus, Charles M.},
	month = jul,
	year = {2023},
	note = {Publisher: American Physical Society},
	pages = {L020505},
}

@article{banszerus_hybrid_2024,
	title = {Hybrid Josephson Rhombus: A Superconducting Element with Tailored Current-Phase Relation},
	author = {Banszerus, L. and Andersson, C. W. and Marshall, W. and Lindemann, T. and Manfra, M. J. and Marcus, C. M. and Vaitiek\ifmmode \dot{e}\else \.{e}\fi{}nas, S.},
	journal = {Phys. Rev. X},
	volume = {15},
	issue = {1},
	pages = {011021},
	numpages = {12},
	year = {2025},
	month = {Feb},
	publisher = {American Physical Society},
	doi = {10.1103/PhysRevX.15.011021},
	url = {https://link.aps.org/doi/10.1103/PhysRevX.15.011021}
}

@article{banszerus_voltage-controlled_2024,
	title = {Voltage-{Controlled} {Synthesis} of {Higher} {Harmonics} in {Hybrid} {Josephson} {Junction} {Circuits}},
	volume = {133},
	url = {https://link.aps.org/doi/10.1103/PhysRevLett.133.186303},
	doi = {10.1103/PhysRevLett.133.186303},
	number = {18},
	urldate = {2025-01-06},
	journal = {Physical Review Letters},
	author = {Banszerus, L. and Marshall, W. and Andersson, C. W. and Lindemann, T. and Manfra, M. J. and Marcus, C. M. and Vaitiekėnas, S.},
	month = oct,
	year = {2024},
	note = {Publisher: American Physical Society},
	pages = {186303},
}

@article{zheng_coherent_2024,
	title = {Coherent {Control} of a {Few}-{Channel} {Hole} {Type} {Gatemon} {Qubit}},
	volume = {24},
	issn = {1530-6984},
	url = {https://doi.org/10.1021/acs.nanolett.4c00770},
	doi = {10.1021/acs.nanolett.4c00770},
	number = {24},
	urldate = {2024-08-29},
	journal = {Nano Letters},
	author = {Zheng, Han and Cheung, Luk Yi and Sangwan, Nikunj and Kononov, Artem and Haller, Roy and Ridderbos, Joost and Ciaccia, Carlo and Ungerer, Jann Hinnerk and Li, Ang and Bakkers, Erik P.A.M. and Baumgartner, Andreas and Schönenberger, Christian},
	month = jun,
	year = {2024},
	note = {Publisher: American Chemical Society},
	pages = {7173--7179},
}

@article{zhuo_hole-type_2023,
	title = {Hole-type superconducting gatemon qubit based on {Ge}/{Si} core/shell nanowires},
	volume = {9},
	copyright = {2023 The Author(s)},
	issn = {2056-6387},
	url = {https://www.nature.com/articles/s41534-023-00721-9},
	doi = {10.1038/s41534-023-00721-9},
	number = {1},
	urldate = {2024-08-29},
	journal = {npj Quantum Information},
	author = {Zhuo, Enna and Lyu, Zhaozheng and Sun, Xiaopei and Li, Ang and Li, Bing and Ji, Zhongqing and Fan, Jie and Bakkers, E. P. a. M. and Han, Xiaodong and Song, Xiaohui and Qu, Fanming and Liu, Guangtong and Shen, Jie and Lu, Li},
	month = may,
	year = {2023},
	note = {Publisher: Nature Publishing Group},
	pages = {1--7},
}

@article{larsen_semiconductor-nanowire-based_2015,
	title = {Semiconductor-{Nanowire}-{Based} {Superconducting} {Qubit}},
	volume = {115},
	url = {https://link.aps.org/doi/10.1103/PhysRevLett.115.127001},
	doi = {10.1103/PhysRevLett.115.127001},
	number = {12},
	urldate = {2024-08-28},
	journal = {Physical Review Letters},
	author = {Larsen, T. W. and Petersson, K. D. and Kuemmeth, F. and Jespersen, T. S. and Krogstrup, P. and Nygård, J. and Marcus, C. M.},
	month = sep,
	year = {2015},
	note = {Publisher: American Physical Society},
	pages = {127001},
}

@article{casparis_superconducting_2018,
	title = {Superconducting gatemon qubit based on a proximitized two-dimensional electron gas},
	volume = {13},
	copyright = {2018 The Author(s)},
	issn = {1748-3395},
	url = {https://www.nature.com/articles/s41565-018-0207-y},
	doi = {10.1038/s41565-018-0207-y},
	language = {en},
	number = {10},
	urldate = {2024-08-28},
	journal = {Nature Nanotechnology},
	author = {Casparis, Lucas and Connolly, Malcolm R. and Kjaergaard, Morten and Pearson, Natalie J. and Kringhøj, Anders and Larsen, Thorvald W. and Kuemmeth, Ferdinand and Wang, Tiantian and Thomas, Candice and Gronin, Sergei and Gardner, Geoffrey C. and Manfra, Michael J. and Marcus, Charles M. and Petersson, Karl D.},
	month = oct,
	year = {2018},
	note = {Publisher: Nature Publishing Group},
	pages = {915--919},
}

@article{mayer_gate_2020,
	title = {Gate controlled anomalous phase shift in {Al}/{InAs} {Josephson} junctions},
	volume = {11},
	copyright = {2020 The Author(s)},
	issn = {2041-1723},
	url = {https://www.nature.com/articles/s41467-019-14094-1},
	doi = {10.1038/s41467-019-14094-1},
	number = {1},
	urldate = {2024-08-27},
	journal = {Nature Communications},
	author = {Mayer, William and Dartiailh, Matthieu C. and Yuan, Joseph and Wickramasinghe, Kaushini S. and Rossi, Enrico and Shabani, Javad},
	month = jan,
	year = {2020},
	note = {Publisher: Nature Publishing Group},
	pages = {212},
}

@article{larsen_parity-protected_2020,
	title = {Parity-{Protected} {Superconductor}-{Semiconductor} {Qubit}},
	volume = {125},
	url = {https://link.aps.org/doi/10.1103/PhysRevLett.125.056801},
	doi = {10.1103/PhysRevLett.125.056801},
	number = {5},
	urldate = {2024-08-27},
	journal = {Physical Review Letters},
	author = {Larsen, T. W. and Gershenson, M. E. and Casparis, L. and Kringhøj, A. and Pearson, N. J. and McNeil, R. P. G. and Kuemmeth, F. and Krogstrup, P. and Petersson, K. D. and Marcus, C. M.},
	month = jul,
	year = {2020},
	note = {Publisher: American Physical Society},
	pages = {056801},
}

@article{Costas_2023,
  title = {Microscopic study of the Josephson supercurrent diode effect in Josephson junctions based on two-dimensional electron gas},
  author = {Costa, Andreas and Fabian, Jaroslav and Kochan, Denis},
  journal = {Phys. Rev. B},
  volume = {108},
  issue = {5},
  pages = {054522},
  numpages = {13},
  year = {2023},
  month = {Aug},
  publisher = {American Physical Society},
  doi = {10.1103/PhysRevB.108.054522},
  url = {https://link.aps.org/doi/10.1103/PhysRevB.108.054522}
}

@article{goffman_conduction_2017,
	title = {Conduction channels of an {InAs}-{Al} nanowire {Josephson} weak link},
	volume = {19},
	issn = {1367-2630},
	url = {https://doi.org/10.1088/1367-2630/aa7641},
	doi = {10.1088/1367-2630/aa7641},
	number = {9},
	journal = {New Journal of Physics},
	author = {Goffman, M F and Urbina, C and Pothier, H and Nygård, J and Marcus, C M and Krogstrup, P},
	month = sep,
	year = {2017},
	note = {Publisher: IOP Publishing},
	pages = {092002},
}

@article{volkov_proximity_1993,
	title = {Proximity effect under nonequilibrium conditions in double-barrier superconducting junctions},
	volume = {210},
	issn = {0921-4534},
	url = {https://www.sciencedirect.com/science/article/pii/092145349390005B},
	doi = {https://doi.org/10.1016/0921-4534(93)90005-B},
	number = {1},
	journal = {Physica C: Superconductivity},
	author = {Volkov, A. F. and Zaitsev, A. V. and Klapwijk, T. M.},
	year = {1993},
	pages = {21--34},
}

\section*{Acknowledgements}
This work has received funding from ARO NextNEQST (contract no. W911-NF22-10048) program. Javad Shabani acknowledges support from ONR MURI (award no. N00014-22-1-2764) and ARO NextNEQST grant mentioned above.

\section*{Author Contributions}
S.L. and A.B. fabricated the devices, performed the measurements, analyzed and interpreted the data and simulated the theoretical model. M.V. provided theoretical support in simulating the model. J.I., I.L. and J.S. have grown the InAs/Al 2DEG material. V.M. helped in interpreting the data. A.B. and S.L. wrote the paper with inputs from all the authors. V.M. initiated and supervised the project. All authors discussed the results and contributed to the manuscript.   

\section*{Competing Interests}
The authors declare no competing interests.

\end{document}


\maketitle

\newpage
\section{Experimental setup}
\begin{figure*}[h]
    \centering
    \includegraphics[width=0.8\linewidth]{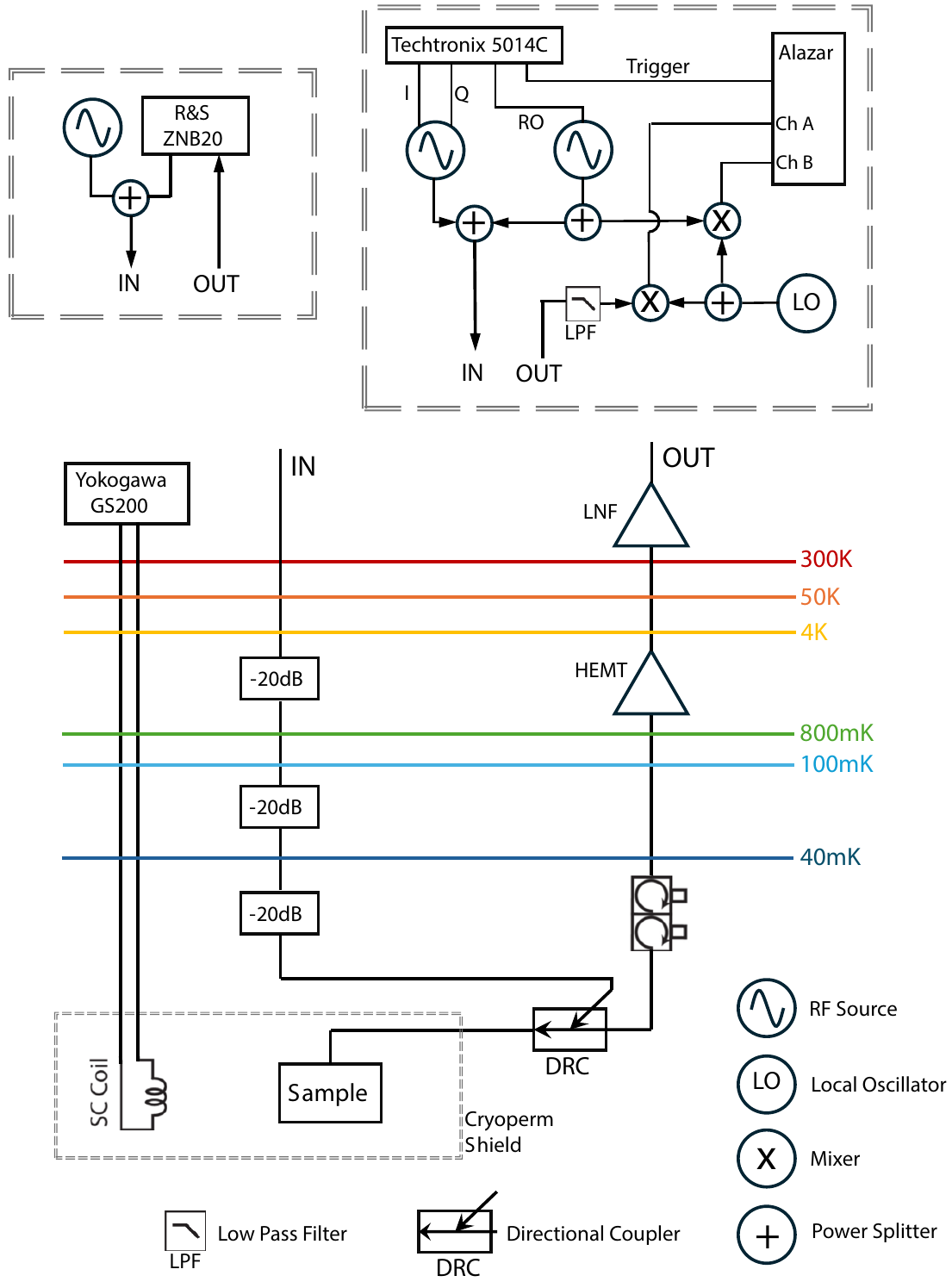} 
    \caption{Schematics of the experimental setup use for measuring the flux-tunable gatemon qubit.}
    \label{suppfig:measurement_schematics}
\end{figure*}

\newpage
\section{One-Tone Spectroscopy}
\begin{figure*}[h]
    \centering
    \includegraphics[width=0.55\linewidth]{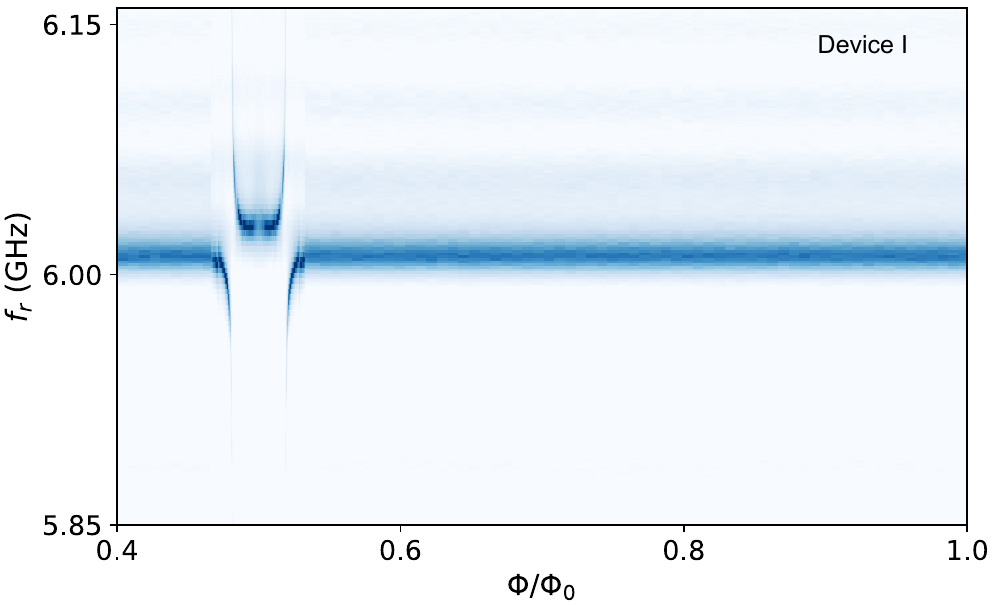} 
    \includegraphics[width=0.55\linewidth]{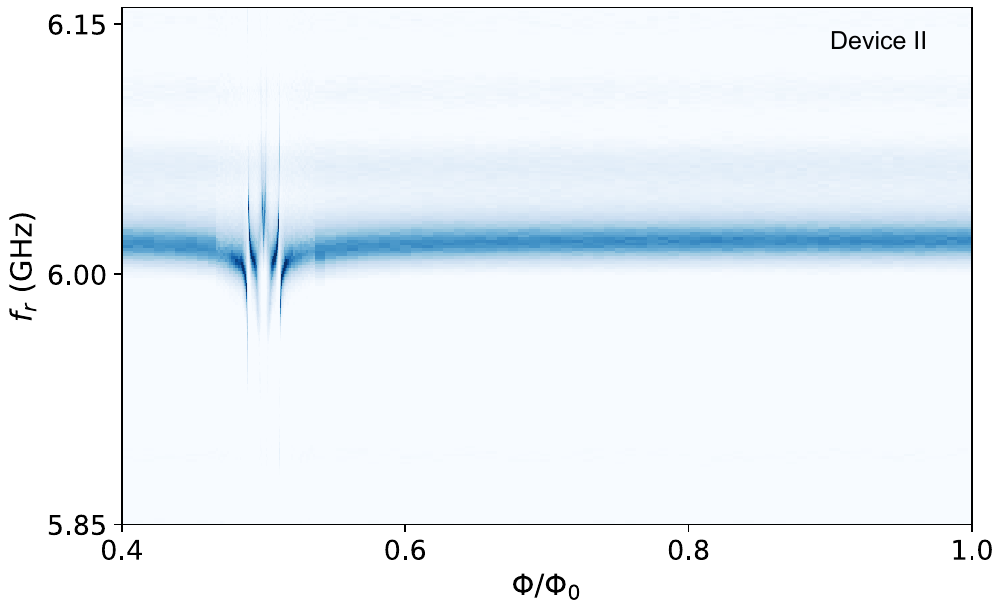} 
    \includegraphics[width=0.55\linewidth]{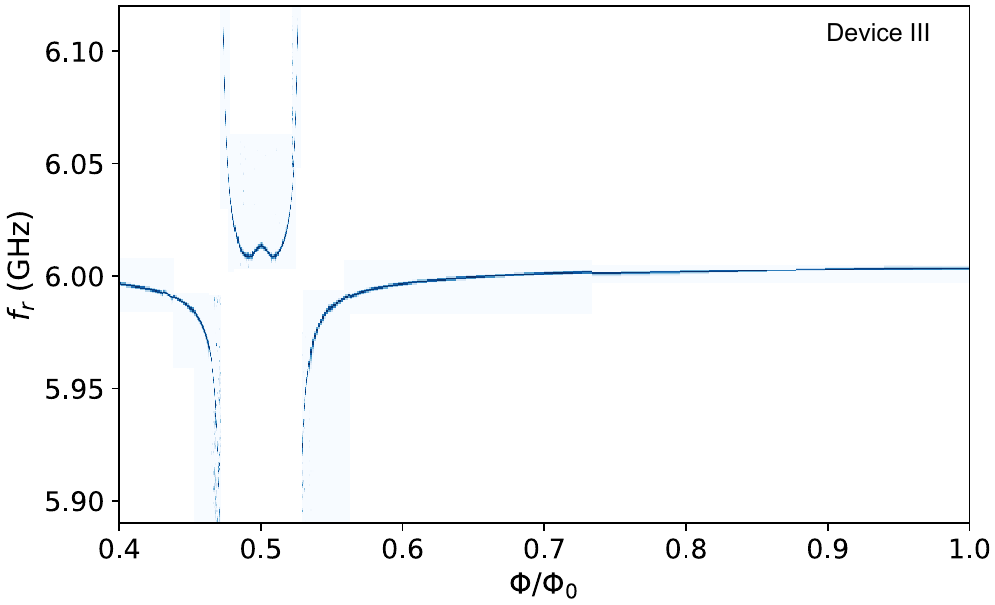} 
    \caption{One-tone spectroscopy showing $|S_{11}|$ as a function of the resonator drive frequency $f_{\textrm{r}}$ and the applied external magnetic flux $\Phi/\Phi_{0}$ for device I, II and III, respectively. The resonator response exhibits a vacuum Rabi splitting with a cavity-qubit coupling strength of $g = 122\,$MHz for device I, $g = 101\,$MHz for device II and $g = 170\,$MHz for device III, respectively.}
    \label{suppfig:Onetone}
\end{figure*}

\newpage
\section{First-order flux insensitivity at $\Phi=0.5\,\Phi_{0}$}
\begin{figure*}[h]
    \centering
    \includegraphics[width=0.495\linewidth]{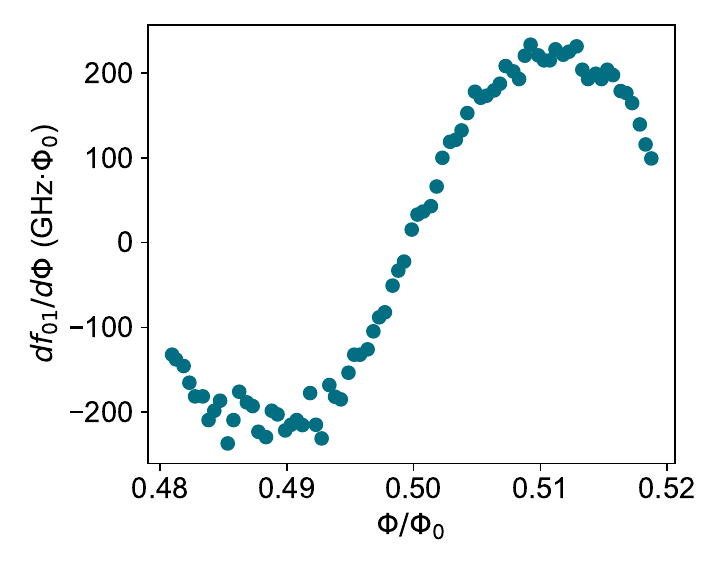} 
    \includegraphics[width=0.495\linewidth]{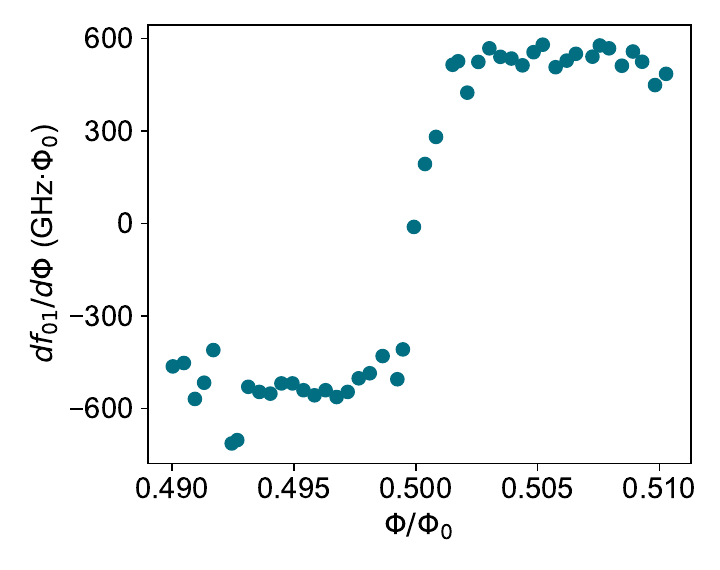} 
    \caption{Derivative of the $\ket{0}-\ket{1}$ transition qubit frequency $\textrm{d}f_{01}/d\Phi$ as a function of the applied external magnetic flux $\Phi/\Phi_{0}$ for device I (left) and device II (right), indicating a first-order insensitivity to flux at the half-flux quanta $\Phi=0.5\,\Phi_{0}$.}
    \label{suppfig:fluxsweetspot}
\end{figure*}

\section{Charge-matrix elements}
\begin{figure*}[h]
    \centering
    \includegraphics[width=0.495\linewidth]{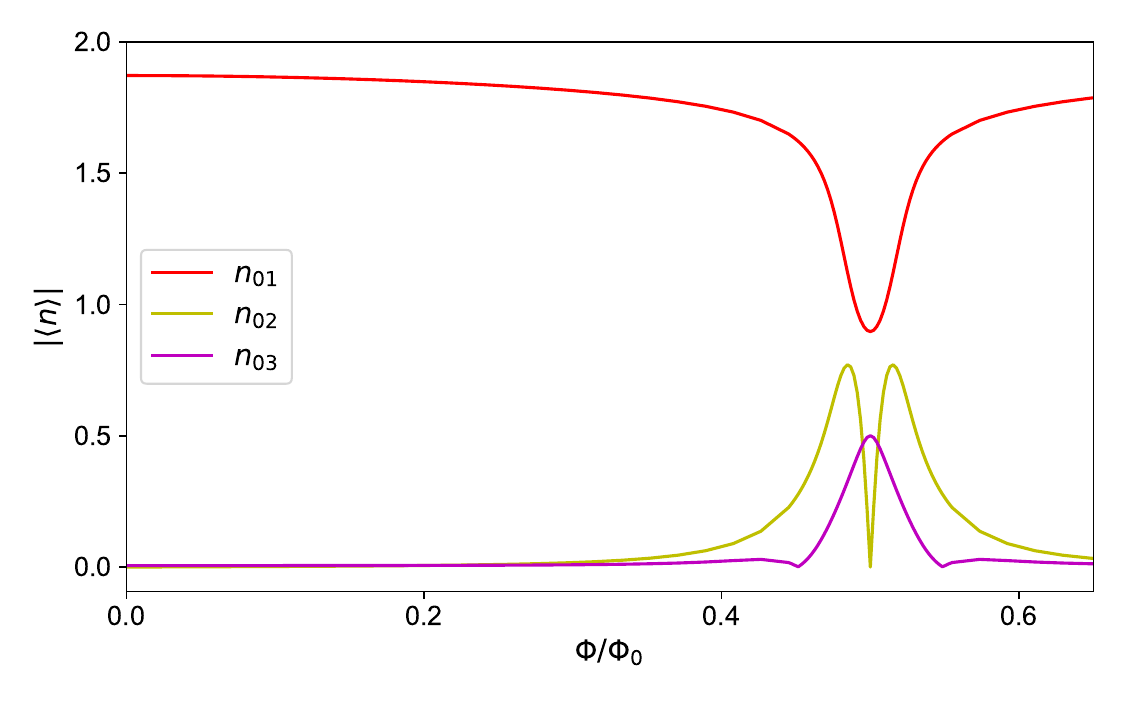} 
    \includegraphics[width=0.495\linewidth]{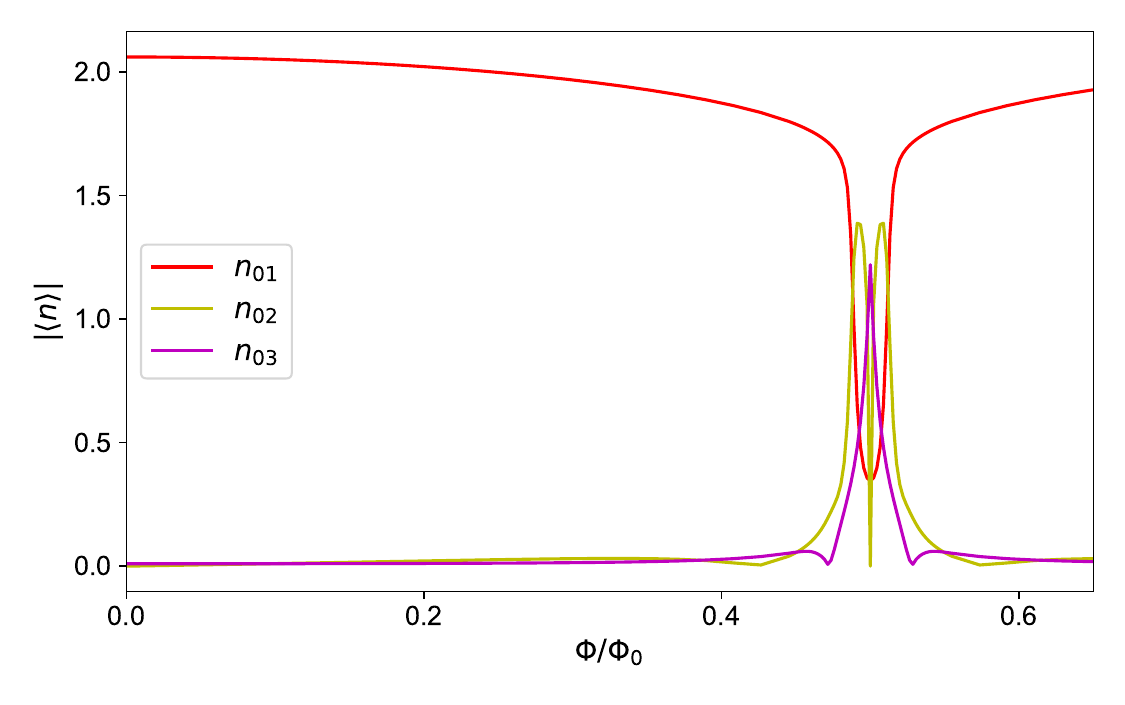}  
    \caption{Charge matrix elements vs flux $\Phi/\Phi_{0}$ for device I (left) and device II (right) calculated using the single-characteristic-transparency model fit of the device spectra.}
    \label{suppfig:chargematrixelements}
\end{figure*}

\newpage
\section{Higher-harmonic model comparison}
\begin{figure*}[h]
    \centering
    \includegraphics[width=0.6\linewidth]{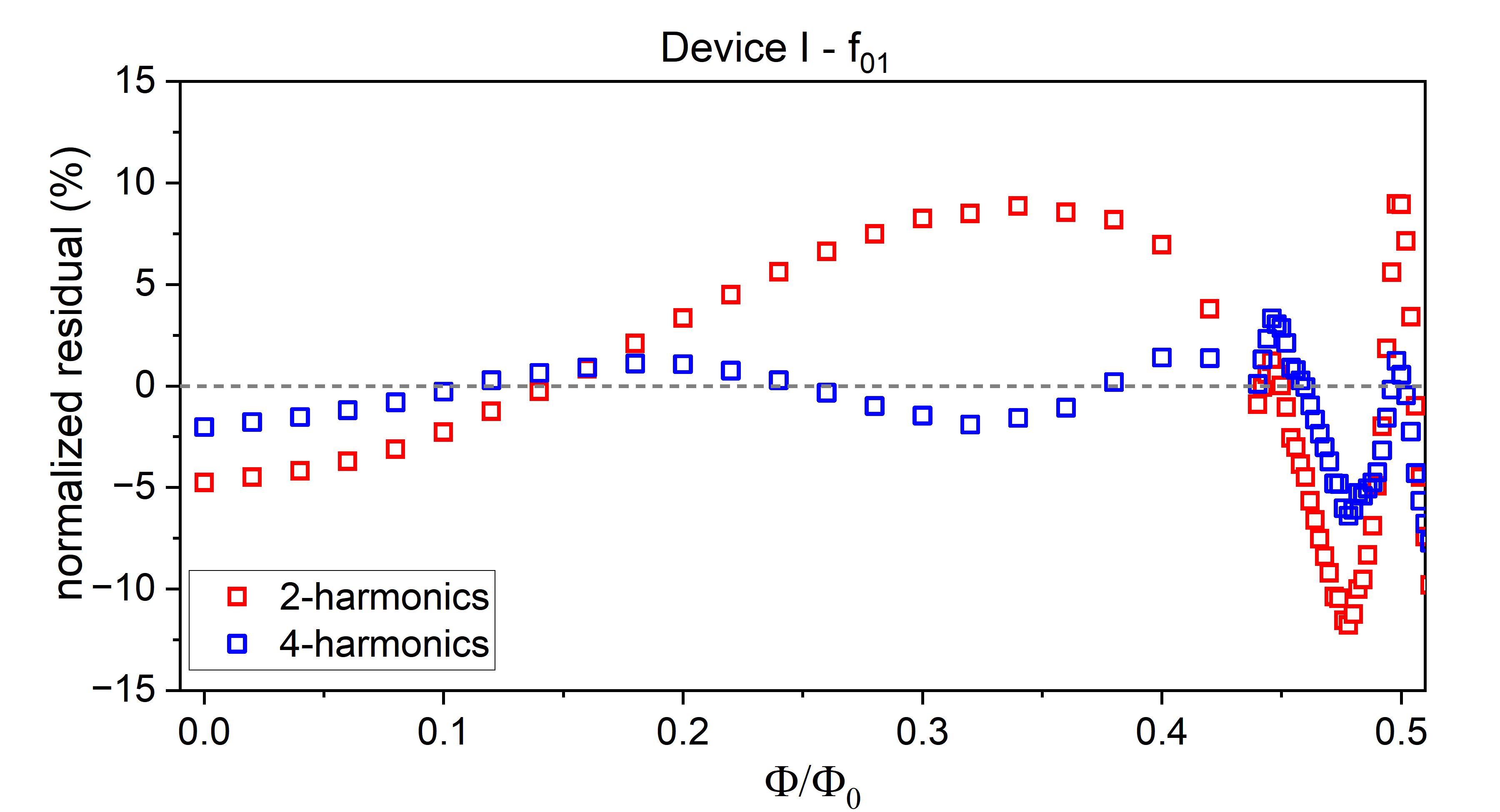} 
    \includegraphics[width=0.6\linewidth]{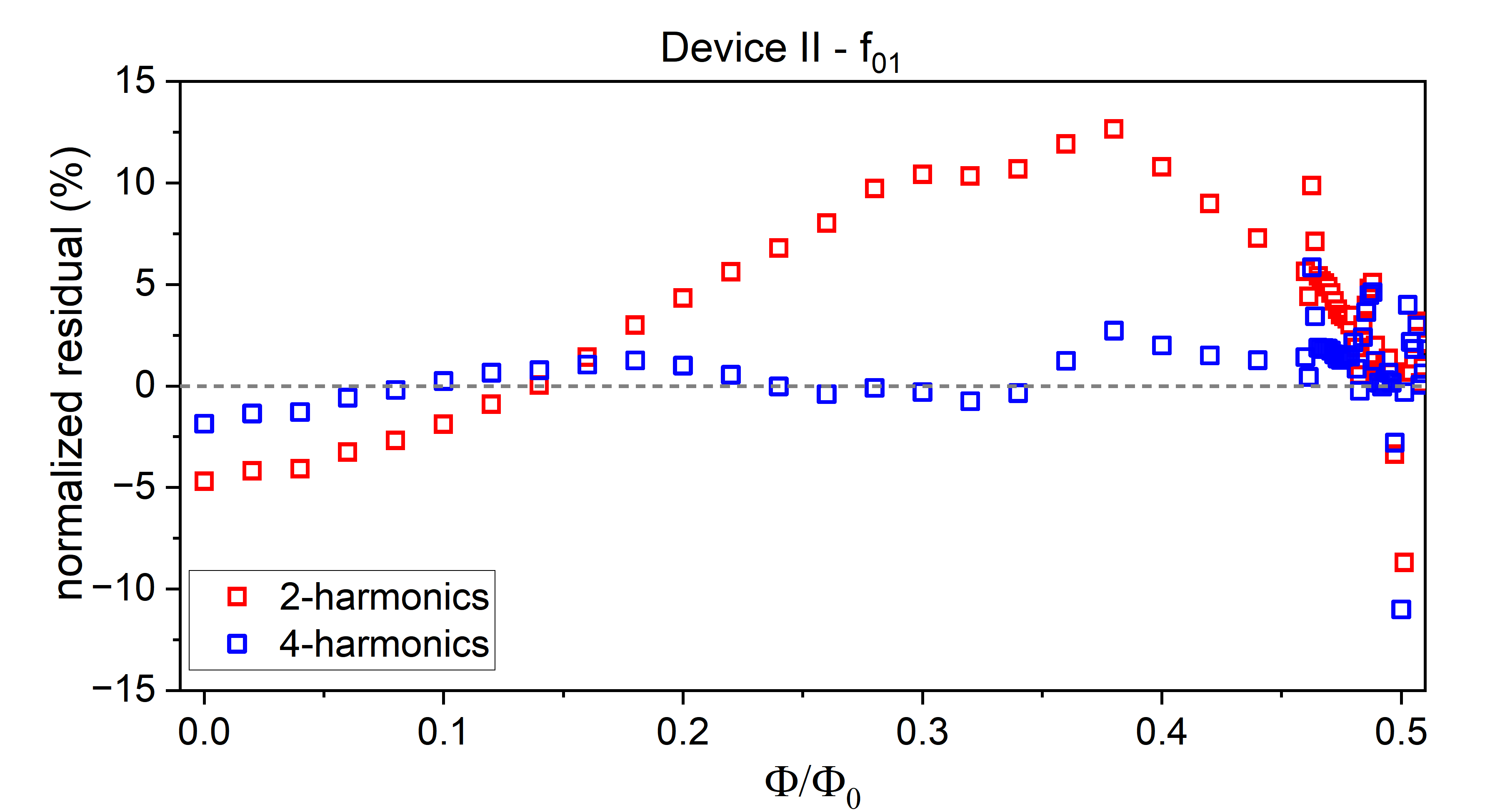}
    \includegraphics[width=0.6\linewidth]{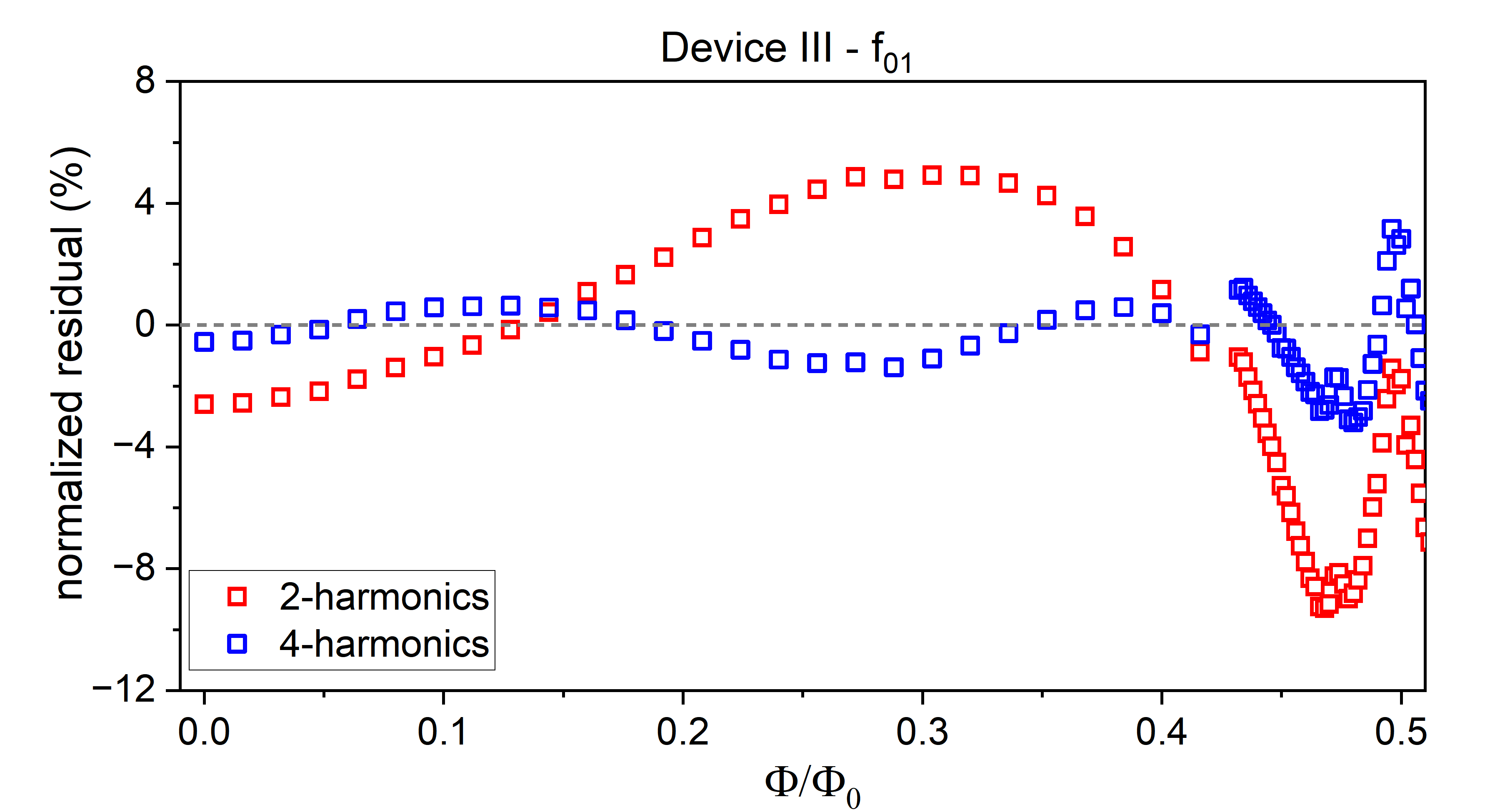}
    \caption{Normalized residual $(f_{\textrm{model}} - f_{\textrm{meas}})/f_{\textrm{meas}}$ as a function of $\Phi/\Phi_{0}$ for the $\ket{0}-\ket{1}$ qubit transition using the higher-harmonic model for device I, II and III, respectively. Here, $k$ refers to the number of leading Fourier harmonic terms used to approximate the Josephson potential energy in $U = \sum_{k} E_{\textrm{J}}^{k} \cos{(k\varphi)}$.}
    \label{suppfig:harmonicmodelcomparison}
\end{figure*}

\newpage
\section{Multi-transparency model}
\begin{figure*}[h]
    \centering
    \includegraphics[width=0.6\linewidth]{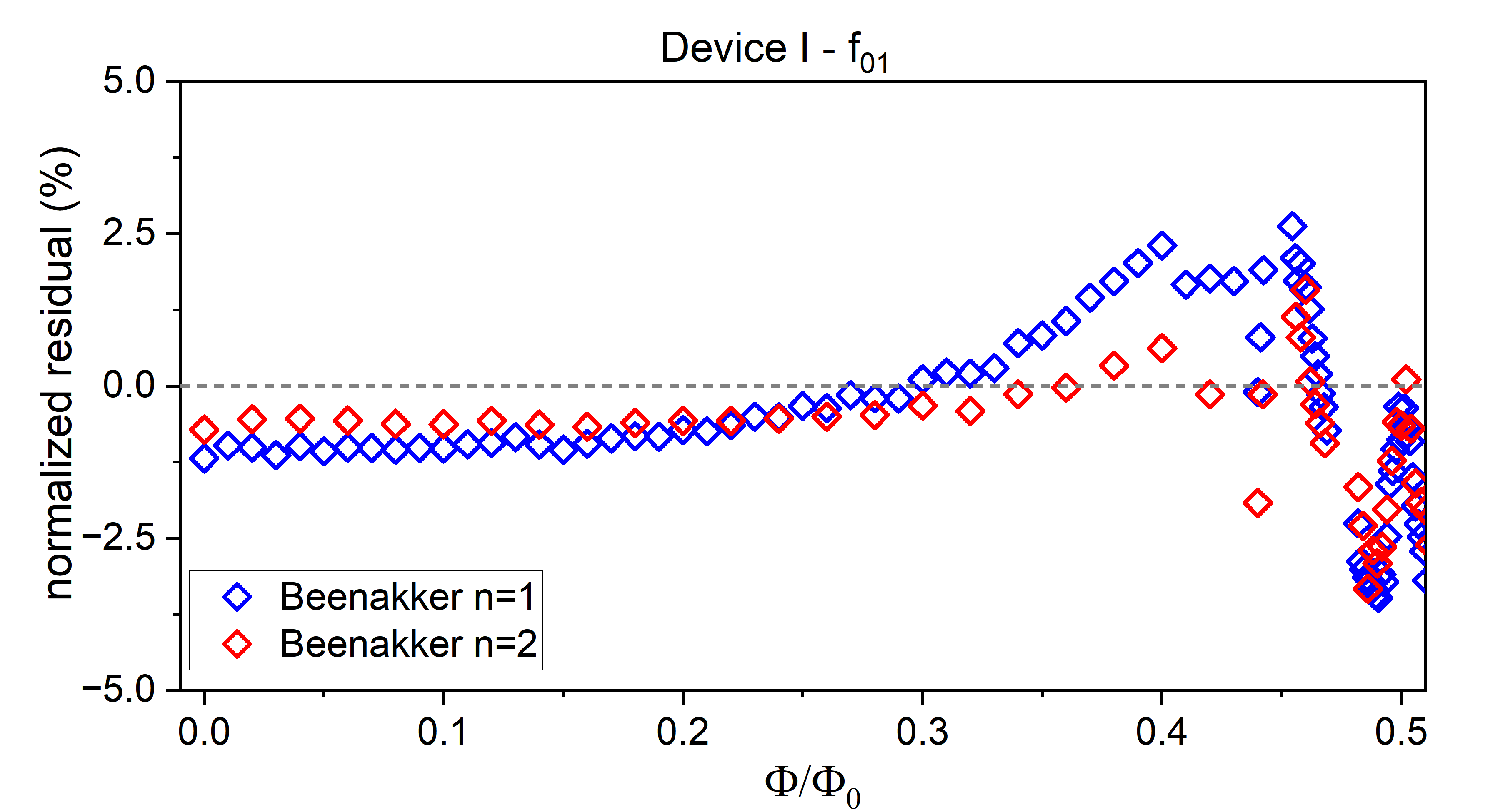} 
    \includegraphics[width=0.6\linewidth]{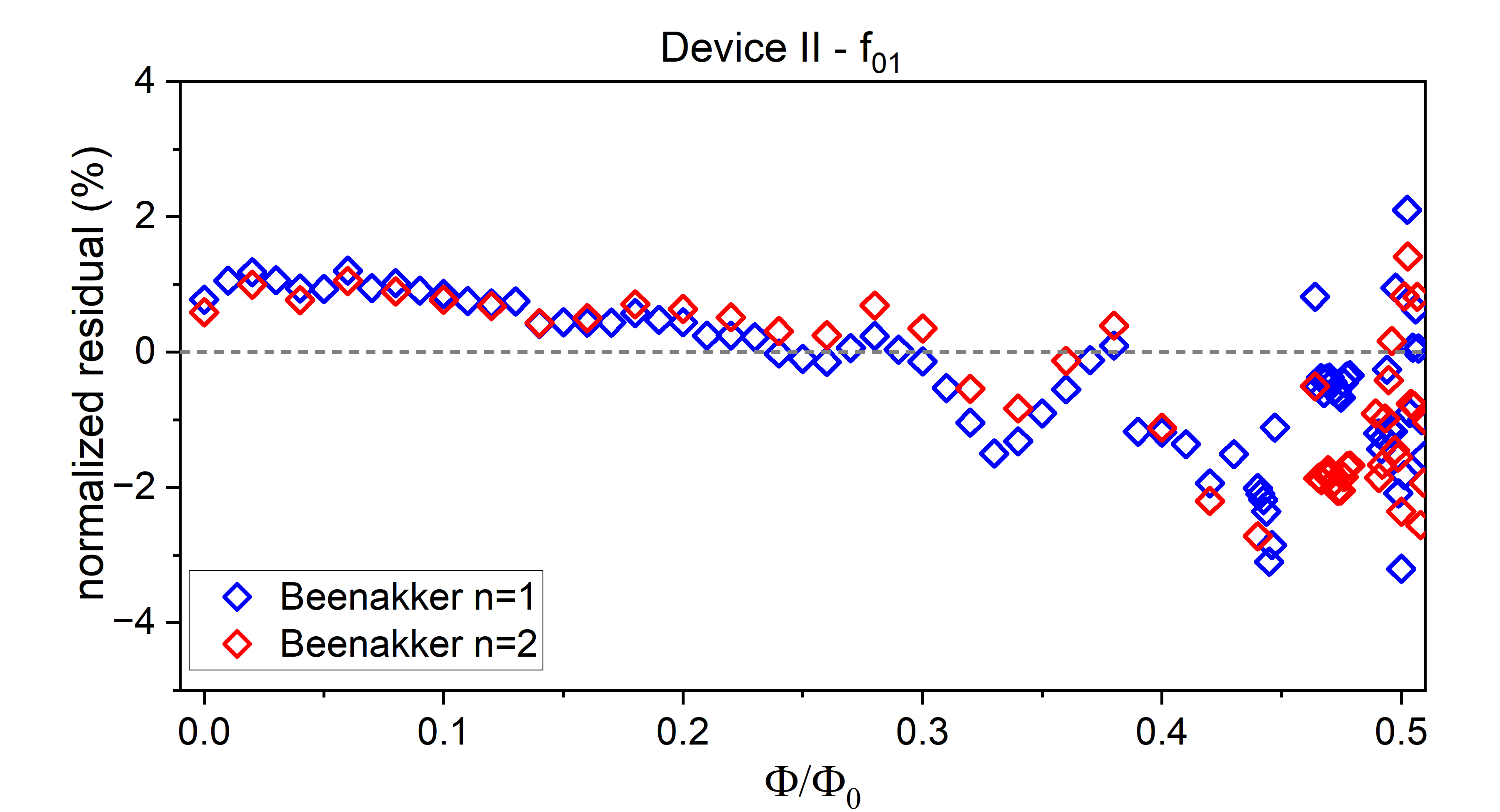}
    \includegraphics[width=0.6\linewidth]{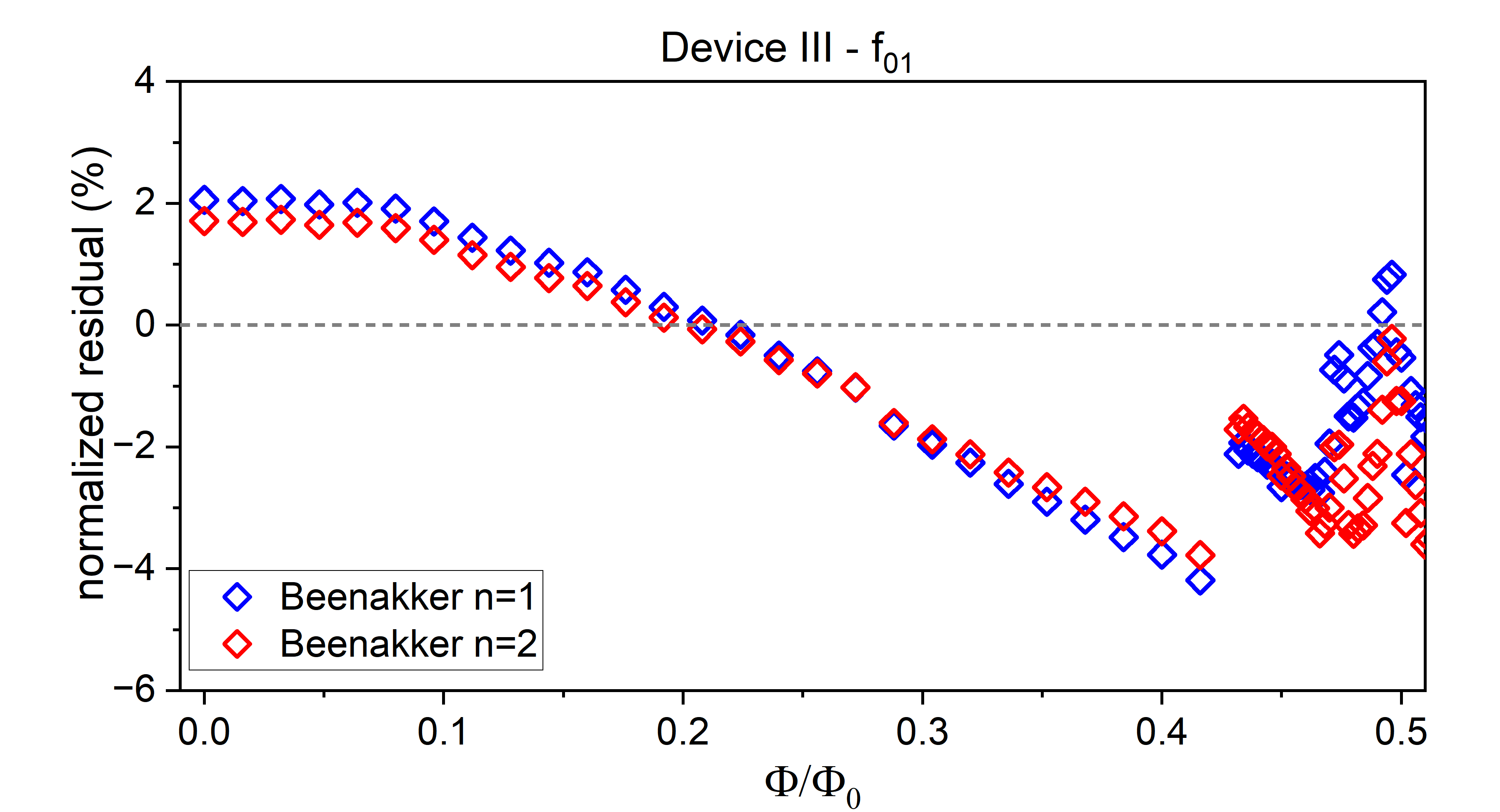}
    \caption{Normalized residual $(f_{\textrm{model}} - f_{\textrm{meas}})/f_{\textrm{meas}}$ as a function of $\Phi/\Phi_{0}$ for the $\ket{0}-\ket{1}$ qubit transition using the single and multiple transparency model for device I, II and III, respectively. Here, $n$ refers to the number of characteristic channel transparencies for each Josephson junction. We observe that increasing $n$ from a single ($n=1$) characteristic channel transparency to two ($n=2$) channel transparencies does not lead to any significant reduction in the residuals.}
    \label{suppfig:multitransparencymodel}
\end{figure*}

\newpage
\section{$T_{1}$ and $T_{2}$ flux dependence}
\begin{figure*}[h]
    \centering
    \includegraphics[width=0.9\linewidth]{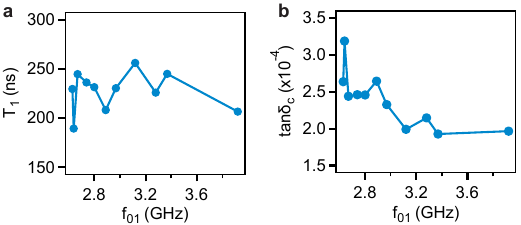} 
    \includegraphics[width=0.9\linewidth]{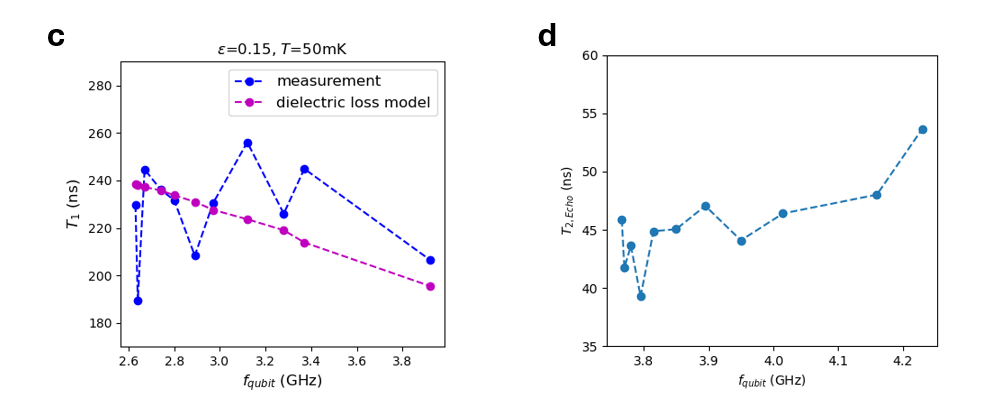}
    \caption{\textbf{a} $T_{1}$ as a function of the $\ket{0}-\ket{1}$ transition qubit frequency $f_{01}$ in the vicinity of half-flux quanta $\Phi = 0.5\,\Phi_{0}$. \textbf{b} The effective dielectric loss tangent $\tan\delta_{\textrm{c}}$ vs the qubit frequency $f_{01}$ extracted from (\textbf{a}) using $\tan\delta_{\textrm{c}} = \frac{1}{T_{1}\omega_{01}}$, where $\omega_{01} = 2\pi f_{01}$. \textbf{c} $T_1$ as a function of the qubit frequency $f_{qubit}$ around half-flux quanta $\Phi = 0.5\,\Phi_{0}$. The blue curve denotes the experimentally measured $T_1$, while the red curve denotes $T_1$ extracted from a dielectric-loss model. This model assumes a weak frequency dependence of the dielectric loss tangent given by: $\frac{\tan\delta(\omega)}{\tan\delta(\omega = 2 \pi \times 2.63\,\textrm{GHz})}= (\frac{\omega}{2 \pi \times 2.63\,\textrm{GHz}})^\varepsilon$ with $\varepsilon=0.15$. The reference value $\tan\delta(\omega = 2 \pi \times 2.63\,\textrm{GHz}) = 2.44 \times 10^{-4}$ is typical for InP dielectric losses. \textbf{d} $T_{2,Echo}$ as a function of the qubit frequency $f_{qubit}$ near $\Phi/\Phi_0 = 0.5$, measured in a separate cooldown, where the qubit frequency at $\Phi/\Phi_0 = 0.5$ is $3.76\,$GHz.}
    \label{suppfig:T1fluxdependence}
\end{figure*}

\newpage
\section{Vacuum Rabi Oscillation Fitting}
\begin{figure*}[h]
    \centering
    \includegraphics[width=0.98\linewidth]{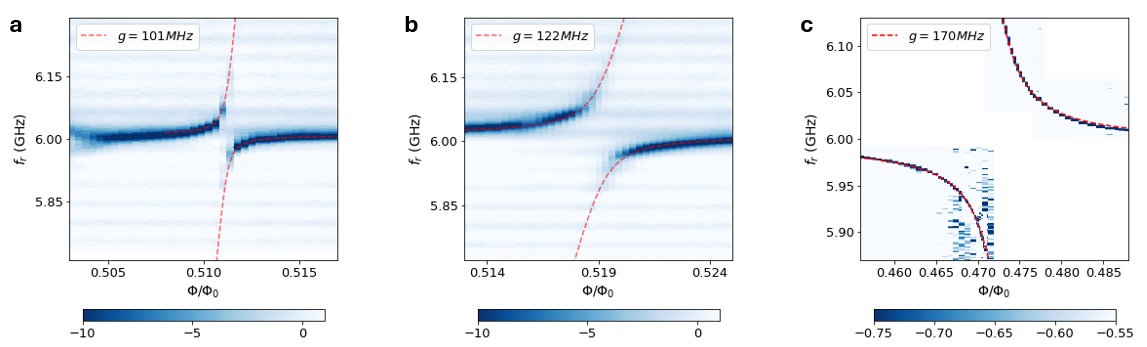}
    \caption{\textbf{a-c} $|S_{11}|$ as a function of flux $\Phi/\Phi_{0}$ and resonator frequency $f_{\textrm{r}}$ for device I, II and III, respectively. The red dashed lines show fits for the vacuum Rabi oscillation to extract the qubit-resonator coupling strength $g$. We note that $|S_{11}|$ is plotted in dB for device I (\textbf{a}) and device II (\textbf{b}), while figure (\textbf{c}) shows the normalized $|S_{11}|$ for device III.}
    \label{suppfig:gfit}
\end{figure*}

In our work, we use the Jaynes-Cummings Hamiltonian denoted by: 
$$H=\hbar\omega_{\ r}a^\dag a+\frac{\hbar\omega_{ge}}{2}\sigma_z+\hbar g\left(a^\dag\sigma_-+a\sigma_+\right)$$
where $a^\dag$,$a$ are the photon creation and annihilation operators, $\sigma_z,\sigma_\pm$ are the Pauli and ladder operators for the qubit, g is the vacuum coupling strength and  $\omega_{\ r},\ \omega_{\ ge}$ are the resonator and the qubit transition frequency respectively. The resultant shape of the avoided crossing is then given by:
$$E_{\pm,n}=\hbar n\omega_r\pm\frac{\hbar}{2}\sqrt{\left(4ng^2+\Delta^2\right)}$$
where $\Delta=\omega_r-\omega_{ge}$ denotes the detuning of the qubit from the resonator frequency.

\newpage
\section{Single-Transparency Model Error Analysis}
\begin{figure*}[h]
    \centering
    \includegraphics[width=0.9\linewidth]{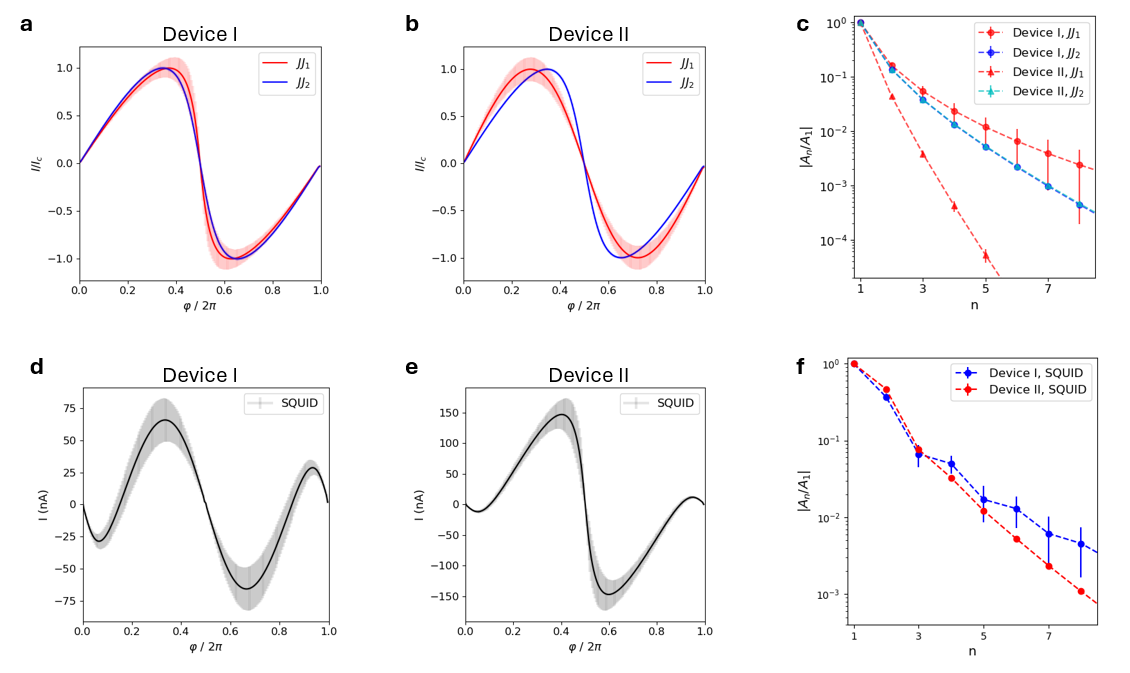}
    \caption{\textbf{a,b} Current-phase relation (CPR) extracted from the single-transparency model for device I and II, with error bands indicating the 95\% confidence intervals. We use the convention where we assign the  notation JJ1 to the junction with larger critical current: $I_{c1}$ = 178.5nA$\pm$19.7nA and $I_{c2}$=71.9nA$\pm$1.1nA for device 1, $I_{c1}$=138.0nA$\pm$17.6nA and $I_{c2}$ = 100.8nA$\pm$1.2nA for device II. \textbf{c} The corresponding normalized harmonic amplitude $|A_{\textrm{n}}/A_{1}|$ and their error bars for each JJ in device I and II. \textbf{d,e} CPR of the SQUID loop at $\Phi/\Phi_0 = 0.5$ with 95\% confidence interval error bands for device I and II. \textbf{f} The corresponding normalized harmonic amplitude $|A_{\textrm{n}}/A_{1}|$ and their error bars for the SQUID loops in device I and II.}
    \label{suppfig:ModelError}
\end{figure*}

We first determine the confidence intervals for the fitting parameters in table 1 by calculating the covariance matrix of the nonlinear least-squares fit. We follow White’s heteroskedasticity-consistent method [H. White, Econometrica, vol. 48, no. 4, pp. 817–838 (1980)], where we first calculate the Jacobian matrix J of the fit numerically using a finite-difference scheme. The covariance matrix C is then approximated as:
$$C=\epsilon^2\left(J^TJ\right)^{-1}$$
where $\epsilon$ is the mean standard error of the fit. The 95$\%$ confidence interval for each fitted parameter $\beta_i$ is then given by:
$$\beta_i\pm Z\sqrt{C_{ii}}$$
where $Z\approx1.96$ is the Z-score that corresponds to the 95$\%$ confidence interval.

To propagate these uncertainties to the extracted CPR and the corresponding Fourier harmonics, we employ a bootstrapping scheme with 500 Monte-Carlo samples. Each parameter $\beta_i$ is assumed to follow a normal distribution with standard deviation $\sigma=\sqrt{C_{i,i}}$, which is justified by the large number of data points relative to the number of fit parameters. Unphysical parameters, such as transmission exceeding unity, are excluded by post-selection. For each valid sample, the CPR is then computed numerically on a uniform grid of 200 phase points, and the harmonic amplitudes are extracted using a Fast Fourier Transform (FFT). The reported error bars for the current and the harmonic amplitude, as shown in supplementary figure~\ref{suppfig:ModelError}, correspond to the median 95$\%$ interval of the resulting distributions.


\newpage
\section{Comparison of charge-matrix elements}
\begin{figure*}[h]
    \centering
    \includegraphics[width=0.60\linewidth]{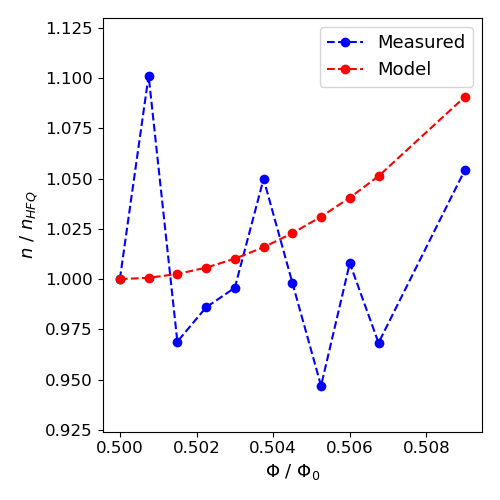} 
    \caption{Normalized charge matrix elements as a function of flux $\Phi/\Phi_{0}$ for device I, where the blue curve is extracted from the measured $T_1$ times in supplementary figure~\ref{suppfig:T1fluxdependence}a using the relation: $1/T_{1} \propto E_{C} |n_{01}|^{2}\tan\delta$, where $E_{C}$ is the charging energy, $n_{01}$ is the charge matrix element and $\tan\delta$ is the dielectric loss tangent. The red curve is calculated using our single-transparency model, }
    \label{suppfig:chargematrixelements}
\end{figure*}

\section{Al coherence length and InAs mean-free path}

The validity of the short-junction approximation is central to our modeling of the Andreev bound states (ABS) and the current-phase relation (CPR). In a superconductor-semiconductor-superconductor (S-Sm-S) junction, the short-junction limit is defined by: 
$$L\ll\ \xi=\frac{\hbar v_{F}}{\pi \Delta}$$ 
where L is the JJ length, $\xi$ is the superconducting coherence length, $v_F$ is the Fermi velocity in the semiconductor and $\Delta$ is the induced superconducting gap. Using $\Delta \approx 230\,\mu$eV determined from the Al critical temperature of $T_{c}\sim 1.5\,$K [Yuan et al., J. Vac. Sci. Technol. A 39, 033407 (2021)] and $v_F\sim 10^{6}\,$m/s corresponding to an electron density of $n_s\approx\ 7.68\ X\ {10}^{11}\textrm{cm}^-2$, we estimate $\xi \sim 1\,\mu$m. For our junction length L=250 nm, this gives $\frac{L}{\xi}\sim 0.25$, placing our devices in the crossover of the short and intermediate regime.

The mean free path of the InAs 2DEG is given by:
$$l=\frac{\hbar k_F\mu}{e}$$
where $k_F=\sqrt{\left(2\pi n_s\right)}$  is the 2D Fermi wavevector, $n_s$ is the electron carrier density and $\mu$ is the electron mobility. Using the measured values of $n_s\approx\ 7.68\ X\ {10}^{11}\,\textrm{cm}^{-2}$ and $\mu\approx20,000\,\textrm{cm}^{2}/Vs$, we obtain $l\approx290$ nm. Since the junction length is $L = 250$ nm, the ratio $l/L\approx1.2$ places our device in the quasi-ballistic regime, but well away from the strongly diffusive limit ($l\ll L$). In this regime, many electron trajectories can traverse the junction without any scattering, giving rise to ballistic contributions to transport, while a non-negligible fraction will scatter, resulting in only a few effective high-transparency channels that dominate supercurrent transport. Although diffusive contributions to transport cannot be entirely excluded, the combination of length scales indicates that our junctions operate in the quasi-ballistic regime, consistent with transport picture presented in the manuscript.




